\documentclass[%
superscriptaddress,
preprint,
 amsmath,amssymb,
 aps,
floatfix,
]{revtex4-2}

\usepackage{graphicx}
\usepackage{dcolumn}
\usepackage{bm}
\usepackage[english]{babel}

\begin{document}

\preprint{APS/123-QED}

\title{Generation of highly mutually coherent hard x-ray pulse pairs with an amplitude-splitting delay line}

\author{Haoyuan Li}
 \altaffiliation[Also at ]{Physics Department, Stanford University, Stanford, California, 94305, U.S.A.}
\author{Yanwen Sun}%
\affiliation{%
 Linac Coherent Light Source, SLAC National Accelerator Laboratory, Menlo Park, California, 94025, U.S.A.
}%
\author{Joan Vila-Comamala}
\affiliation{
 Paul Scherrer Institute, Forschungsstrasse 111, 5232 Villigen PSI,
Switzerland
}%
\author{Takahiro Sato}%
\author{Sanghoon Song}%
\author{Peihao Sun}%
 \altaffiliation[Also at ]{Physics Department, Stanford University, Stanford, California, 94305, U.S.A.}
\author{Matthew H Seaberg}%
\author{Nan Wang}%
 \altaffiliation[Also at ]{Physics Department, Stanford University, Stanford, California, 94305, U.S.A.}
\author{J. B. Hastings}%
\author{Mike Dunne}%
\author{Paul Fuoss}%
\affiliation{%
 Linac Coherent Light Source, SLAC National Accelerator Laboratory, Menlo Park, California, 94025, U.S.A.
}%


\author{Christian David}
\affiliation{
 Paul Scherrer Institute, Forschungsstrasse 111, 5232 Villigen PSI,
Switzerland
}%
\author{Mark Sutton}
\affiliation{%
 Physics Department, McGill University, Quebec, H3A 2T8, Canada
}%

\author{Diling Zhu}
 \email{dlzhu@slac.stanford.edu}
\affiliation{%
 Linac Coherent Light Source, SLAC National Accelerator Laboratory, Menlo Park, California, 94025, U.S.A.
}%

\date{\today}

\begin{abstract}
Beam splitters and delay lines are among the key building blocks of modern-day optical laser technology.
Progress in x-ray free electron laser source development and applications over the past decade is calling for their counterpart operating at the Angstrom wavelength regime.
Recent efforts in x-ray optics development demonstrate relatively stable delay lines that most often adopt the division-of-wavefront approach for the beam splitting and recombination.
However, the two exit beams in such configurations struggle to achieve sufficient mutual coherence to enable applications such as interferometry, correlation spectroscopy, and nonlinear spectroscopy.
We present an experimental realization of the generation of highly mutually coherent pulse pairs using an amplitude-split delay line design based on transmission grating beam splitters and channel-cut crystals.
The performance of the prototype system was analyzed in the context of x-ray coherent scattering and correlation spectroscopy, where nearly identical high-contrast speckle patterns from both branches were observed.
We show in addition the high level of dynamical stability during continuous delay scans, a capability essential for high sensitivity ultra-fast measurements.
\end{abstract}

\maketitle


\section{Introduction}
Modern optical laser technology relies heavily on high performance optical components that enable precise manipulation of the electromagnetic field at sub-wavelength spatial/temporal scales.
The rapid progress in x-ray laser sources over the past decade, in particular in the form of x-ray free electron lasers, has opened up the potential of extending many experimental laser methodologies from optical into atomic scale wavelengths~\cite{seddon2017short}.
The realization of these methodologies, such as interferometry, dynamic light scattering, and nonlinear spectroscopy, requires the development of optical components and systems to manipulate x-ray laser beams in a similar fashion as for optical lasers.
Optics for precision control of femtosecond x-ray pulses in the multi-dimensional space of time, space, spectrum, and propagation direction are highly desired.

Among the basic x-ray optical components, we have seen great advances in mirrors and lenses~\cite{yamauchi2002figuring, yamauchi2011single,seiboth2017perfect}.
On the other hand, we have yet to establish effective beam splitters and delay lines, which are required for multi-beam x-ray laser beam experiments such as x-ray interferometry, x-ray wave mixing, as well as x-ray photon correlation spectroscopy (XPCS)~\cite{mimura2014generation, tanaka2002coherent,schweigert2008double,Grubel2007,Gutt2009}.
X-ray photon correlation spectroscopy, for example, is an extension of dynamic light scattering, reaching the atomic length scale and femtosecond time scale.
It has the potential to directly probe the femtosecond and picosecond time scale (fs-ps) dynamics of disordered matters and their phase transitions that are currently inaccessible by any other existing experimental probes, e.g. many-body dynamics in super-cooled liquids, dynamical heterogeneity, and strong-to-fragile transitions~\cite{wang2019universal,PhysRevLett.113.117801}.
Strong interest in those multi x-ray pulse capabilities have driven tremendous efforts in the design and implementation of hard x-ray split-delay optics at several X-ray FEL facilities over the past decade~\cite{roseker2009performance,osaka2013bragg,lu2018development,rysov2019compact,osaka2016wavelength,zhu2017development,shi2018multi,sun2019compact, sun2019design}.
Existing designs and systems have established routine and stable delivery of hard x-ray pulse pairs with good efficiency in recent years~\cite{osaka2016wavelength,shi2018multi}. 
One last and the most demanding requirement that has yet to be met is the preservation of mutual coherence between the two pulses in a pulse pair.
Two of the primary remaining limitations relate to the performance of the crystal-optics-based beam splitters~\cite{sun2020realizing}, and the pulse front tilt induced by the asymmetric channel-cut crystals that were used to enhance the beam stability and to change the delay time \cite{sun2019compact}.

Coauthors of this paper proposed a new optical design in 2020, which uses transmission gratings as the beam splitter and recombiner, and a dispersion-compensated all-channel-cut 8-bounce delay line to adjust the path length~\cite{li2020design}.
Numerical studies showed a significant performance enhancement.
In this paper, we report the first experimental realization of this novel optical concept. 
This prototype device demonstrates the capability of generating nearly-identical pulse pairs, as manifested in the nearly-identical high contrast speckle patterns obtained from both branches, which is a direct proof of a high degree of mutual coherence.
We show in addition the capability of maintaining this high mutual coherence during continuous delay scans, which is unprecedented and essential for high sensitivity ultra-fast measurements.

\section{Experiment Setup}
\begin{figure*}[bht!]
    \centering
     \includegraphics[width=0.9\textwidth,keepaspectratio]{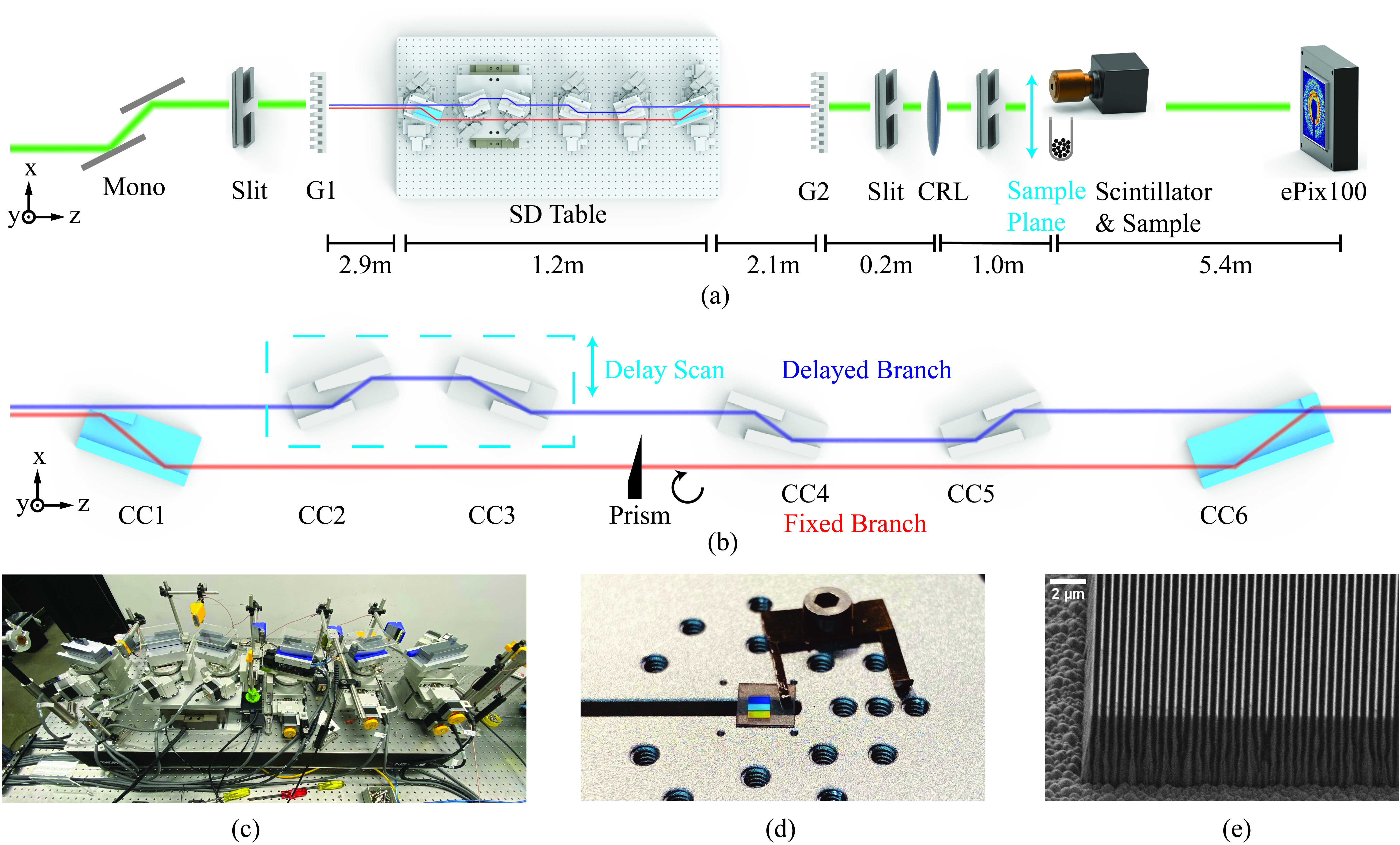}
    \caption{(a) Layout of the experiment setup. The x-ray pulse propagates from left to right, sequentially through, the upstream diamond monochromator (Mono), the $1^{st}$ diamond grating ($G1$), the optical table for the split-delay device (SD table), the $2^{nd}$ diamond grating ($G2$), two JJ-X-ray slits ($Slit$), the compound refractive lens (CRL), the sample or the beam profile monitor assembly (Scintillator \& Sample), and the ePix100 x-ray direct detector (ePix100).
    (b) Schematics and beam paths of channel-cut crystals on the SD table. 
    (c) A photograph of the SD table.
    (d) A photograph of $G1$ with a holder. 
    (e) An electron microscopy image of the grating. }
    \label{fig:schematics_and_photos}
\end{figure*}

The overall layout of the experiment setup is shown in figure~\ref{fig:schematics_and_photos} (a) and (b).
The design and numerical analysis of this setup was described in Ref~\cite{li2020design}.
The experiment was performed at the X-ray Pump Probe instrument using the diamond (111) monochromator which selects a $\sim$0.5~eV bandwidth from the incident FEL output of a boarder bandwidth \cite{Chollet2015, zhu2014performance}. 
Downstream of the monochromator, the size and transverse position of the incident beam was further defined by upstream slits.
The slits were closed down to 500~$\mu$m to form a square aperture.

The x-ray beam was first split by the upstream transmission grating $G1$.
The grating was fabricated on a single crystal diamond substrate of $ 4~mm\times 4~mm \times 100~\mu m$.
The grating pattern was produced by high resolution electron beam lithography in  hydrogen silsesquioxane resist.
The pattern was then transferred to the diamond substrate by oxygen plasma assisted reactive ion etching \cite{Mikata2017, Li2020grating}.
The grating period is 500~nm and the overall size is $0.6~ mm \times 1.5~mm$.
A photograph and a high resolution scanning electron microscopy image of the diamond grating are shown in  figure~\ref{fig:schematics_and_photos} (d) and (e) , respectively. 
The groove depth of the grating is about 5 $\mu m$.
During the experiment, the grating was rotated around the x axis to increase the effective groove depth towards 8 $\mu m$ in order to reach a $\pi$ phase shift that maximizes the photon flux in $\pm1$ diffraction orders. 
Diffracted beams from $G1$ were delivered to the main crystal-optics table (SD table) through an evacuated beam path.
The distance between $G1$ and the first channel-cut crystal ($CC1$) on the SD table was 2.90~m.

The SD table, shown in figure~\ref{fig:schematics_and_photos}(c), supports the motion control mechanisms for 6 silicon channel-cut crystals (CCs), which are referred to as $CC1$ to $CC6$ in this paper.
$CC1$ and $CC6$ are regular channel-cut crystals with pairs of polished parallel optical surfaces, but different gap sizes: 25.15~mm and 25.8~mm respectively.
$CC2$ to $CC5$ are asymmetric channel-cut crystals (ACC) with asymmetry angles of $5^\circ$ as can be seen from figure~\ref{fig:schematics_and_photos}(b).
The dimensions of the ACCs are identical to those described in Ref~\cite{sun2019compact}.
All 6 CCs utilized (220) Bragg reflections.
Effectively two delay lines were formed by the 6 Bragg reflection pairs. the fixed branch ($CC1$ and $CC6$) and the delayed branch ($CC2$ to $CC5$).
In the delayed branch, $CC2$ and $CC3$ were mounted on a single air-bearing stage.
The relative path length between the two branches can be adjusted by moving the air-bearing stage along the $\textbf{x}$ direction as indicated in figure~\ref{fig:schematics_and_photos}(b).

At 9.83~keV, the spatial separation of $\pm 1$ orders of diffraction is $\sim$1.5~mm at $CC1$.
This distance allows full spatial separation of the two diffraction orders.
The $+1$ order of diffraction from $G1$ was picked up by $CC1$, while the $-1$ order of diffraction was picked up by $CC2$.
The $0$ order diffraction is filtered out by CC2, being outside its reflection bandwidth. 
The SD table was enclosed in a helium environment to reduce air absorption.
Downstream of the SD table, the two exit beam paths from $CC5$ and $CC6$ merged at the second grating $G2$ which shares identical parameters as $G1$.

Further downstream of $G2$, only the $-1$ order of diffraction from the fixed branch and $+1$ order from the delayed branch became parallel to the incident beam. The grating-induced angular dispersion was also fully removed by 2 deflections of opposite directions.

A beryllium Compound Refractive Lens (CRL) was used to focus the beam down to about $\sim$1~$\mu$m at the nominal sample plane.
The slits upstream of the CRL were used to define the illuminated area of the lens to reduce the sensitivity of the focal position to upstream beam position change.
The slits downstream of the CRL were used to block the other diffraction orders from $G2$.
At the nominal sample plane, we could insert either a silica nanoparticle powder sample to produce coherent small angle scattering, or a scintillator based beam profile monitor to directly investigate the spatial property of the focused beam.
After another section of evacuated beam path, 5.40~m further downstream of the sample plane, an ePix100 x-ray area detector \cite{Carini2014} was used to collect speckle patterns of the sample.
A beam stop was positioned in front of the the detector to block the direct beam.

\section{Alignment Procedure}\label{Section:Alignment}
\begin{figure}[bht!]
    \centering
    \includegraphics[width=0.48\textwidth,keepaspectratio]{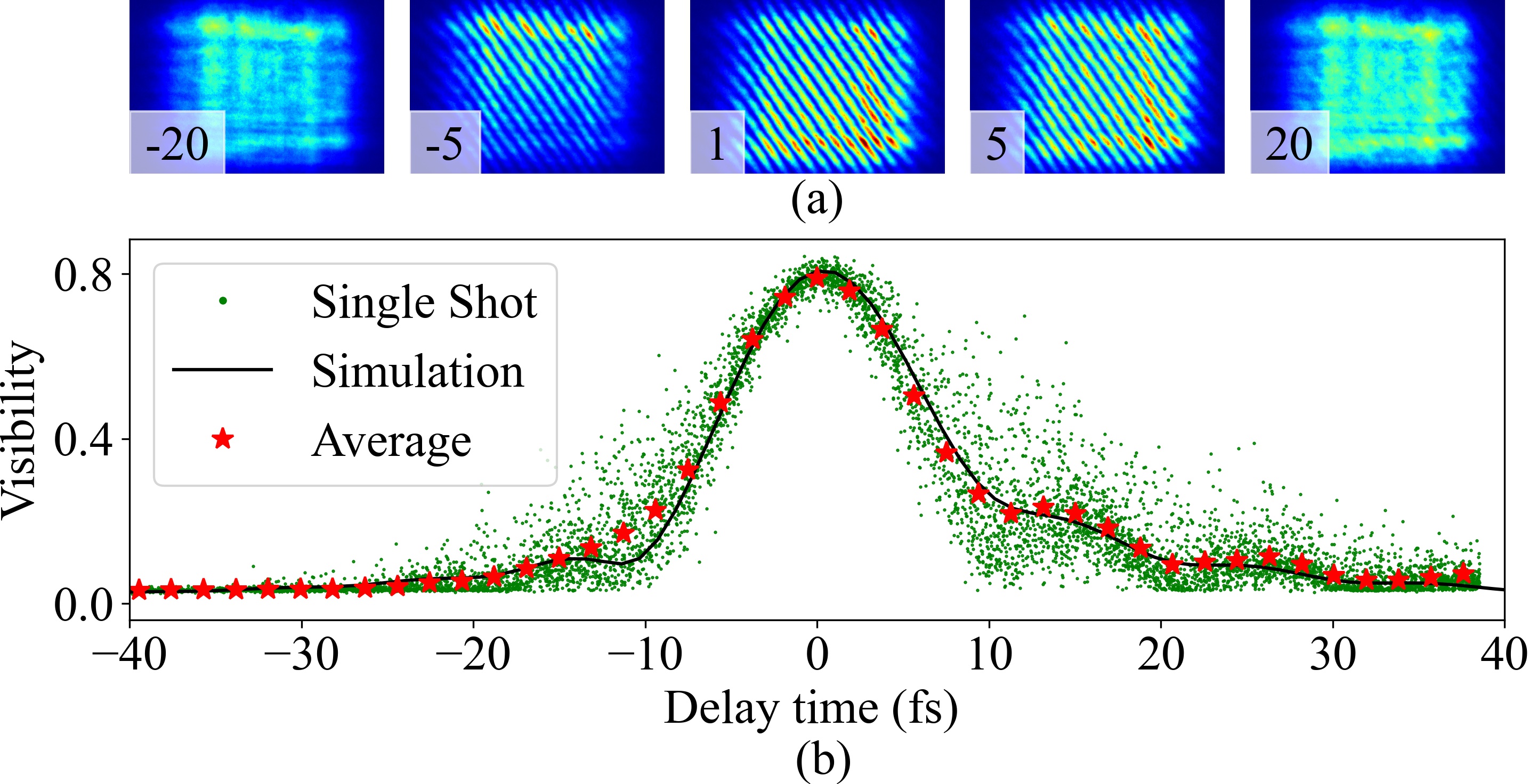}
    \caption{(a) Interference fringes with varying time delay. The number on the lower left corner is the corresponding delay time in fs. (b) Visibility of the interference fringes versus the relative delay time between the two pulses. Green dots are visibility for each single interference pattern, red stars are mean values of single pattern visibility values, and the black curve is the simulation value.}
    \label{fig:T0_and_visibility}
\end{figure}

This section describes the alignment procedure of the system.
First, the position and orientation of $G1$ was optimized by analyzing the forward diffraction with the beam profile monitor at the sample plane.
By adjusting the orientation of $G1$, all diffraction orders were brought to the horizontal plane.
$G1$ was rotated next around the $x$ axis to maximize the flux in the $\pm1$ diffraction orders.
The optimized rotation angle was found to be $72^\circ\pm 2^\circ$.
In the next step, optimal Bragg conditions of $CC1$ and $CC6$ were established using intensity diagnostics behind each crystal.
This was repeated for the delayed branch from $CC2$ to $CC5$.
We then brought the two exit beams from both branches to the same vertical position as the input beam by adjusting the tilting angles of the crystals.
The two beams were then overlapped at the $G2$ location using another scintillator screen.
In addition, an intentional small angular offset was added to the $CC6$ Bragg angle to match the slightly narrower bandwidth of the delayed branch.

To align $G2$, we used the scintillator screen at the sample plane to maximize the $\pm 1$ orders from fixed branch only.
The optimized rotation angle was found to be $73^\circ\pm2^\circ$.
The spatial overlap between the two branches at the sample plane was finally established on the sample plane profile monitor, first with unfocused beams and then with focused beams.

After the two unfocused exit beams were spatially overlapped on the sample plane, a glassy carbon prism was inserted into the fixed branch to create a small crossing angle between the two branches. 
When the two beams overlap with each other within the coherence time, one would observe high contrast interference fringes.
This allows us to determine the \emph{$T_0$} for the split-delay system. 

The steering angle was determined by the prism shape and orientation. 
In our case, this angle was $\sim 5~\mu rad$. 
Detailed analysis is presented in Appendix~\ref{apopendix:prism steering}.
A few examples of interference fringes observed near $T_0$ are shown in figure~\ref{fig:T0_and_visibility} (a).
The relative delay time as a function of the air-bearing stage position \cite{Osaka2017, li2020design} follows the expression:
\begin{equation} \label{equation:general delay formula}
    \Delta(t) = \frac{4\Delta(d)}{c}\frac{\sin{\theta_{Bragg}}\sin{\alpha}}{\sin{(\theta_{Bragg} + \alpha) }},
\end{equation}
where $\Delta(t)$ is the change of the delay time, the $\Delta(d)$ indicates the displacement of the air-bearing stage, the $\theta_{Bragg}$ is the Bragg angle, and $\alpha$ is the asymmetry angle in of the ACC which is $5^\circ$.
The relationship between the visibility of the interference fringes and the delay time allows us to determine $T_0$ and characterize the mutual longitudinal coherence between the two branches.
The fringe visibility is calculated for a large number of single shot patterns at various delays between -40~fs and 40~fs and plotted in figure~ \ref{fig:T0_and_visibility}(b).
The coherence time, i.e. the FWHM of this curve, is determined to be 11.3~fs. 
Since our measurement of $\Delta(d)$ is better than 1~$\mu m$ (the corresponding $\Delta(t)$ is smaller than 1~fs), the uncertainty of the delay time is dominated by this visibility peak. Therefore the accuracy of the delay time is about 10~fs.
A distinguished and asymmetric modulation of the overall bell-shape curve was observed on the tails. 
This modulation is a signature of the temporal tail structure of the output pulses as predicted in Ref~\cite{li2020design}.
The asymmetry can be attributed to the remaining bandwidth and spectral content differences between the two branches.
We are able to reproduce the average contrast as a function of delay time by numerical modeling (the source code is available at the code base \cite{code}).
A prism steering angle of 5~$\mu rad$ and a delayed branch detuning of by 5.2~$\mu rad$ was used in the simulation. 

\begin{figure}[bht!]
    \centering
    \includegraphics[width=0.48\textwidth,keepaspectratio]{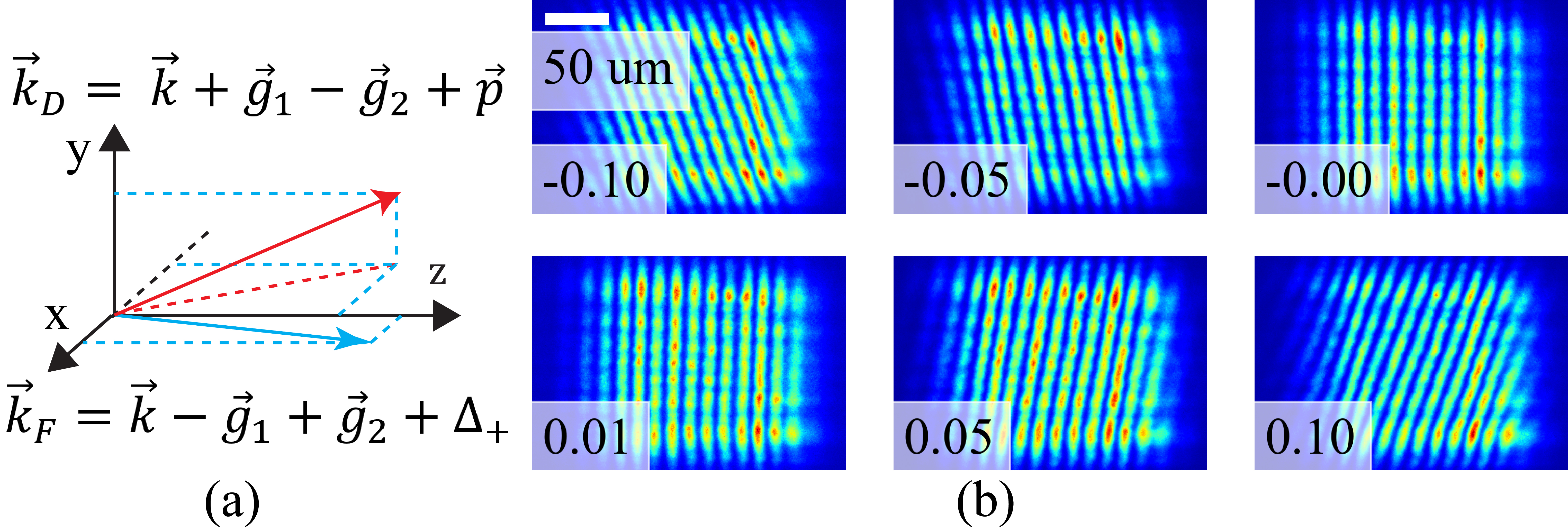}
    \caption{(a) The vector relation between wave vectors from fixed and delayed branch. The $\Vec{k}_D$ represents the output wave vector of the delayed branch. The $\Vec{k}_{F}$ represents the output wave vector of the fixed branch. The $\Vec{k}$ is the incident wave vector, $\Vec{g}_1$ is the photon momentum transfer from $G1$ to the fixed branch, $\Vec{g}_2$ is the photon momentum transfer from $G2$ to the delayed branch, $\Vec{p}$ is the photon momentum transfer from the prism to the fixed branch and $\Delta_+$ is the photon momentum transfer from asymmetric Bragg reflections. We assume that $|\Vec{g}_1| =|\Vec{g}_2|$. Therefore, when $G1$ and $G2$ are perfectly aligned with each other, and $\Vec{p}$ and $\Delta_+$ are in the x-z plane, the interference fringe should be parallel to the $y$ axis.  (b) Interference fringe observed with the prism inserted into the fixed branch. The value on the lower left corners of the images indicate the deviation of the roll angle of $G2$ from the optimal value in degree.}
    \label{fig:fringe_tilt}
\end{figure}

The high contrast interference fringe pattern can be used for evaluating and optimizing the collinearity of the two beams by examining the fringe spacing and orientation.
The prism steers the beam predominantly in the horizontal plane.
Interference fringes parallel to the vertical direction are expected if the two branches are exactly collinear before the prism insertion.
The fringe tilt as shown in figure~\ref{fig:T0_and_visibility} indicates a vertical crossing angle between the two pulses.
As shown in figure~\ref{fig:fringe_tilt}, when the net momentum transfer from $G1$ and $G2$ is not in the horizontal plane, the interference fringe will be tilted with respect to the vertical axis.
By adjusting the orientation of $G2$, we could eliminate the interference fringe tilt angle, thus eliminate the vertical crossing angle.

Quantitatively measuring the fringe spacing also allows optimization of the crossing angle in the horizontal direction. Based on the prism steering calculation, we expected a $5.0~\mu rad$ crossing between the two branches if they were parallel prior to prism insertion (see detailed calculation in Appendix~\ref{apopendix:prism steering}). The measured fringe period shown here indicated that the crossing angle of the two branches was 10~$\mu rad$.
The excess of horizontal angular crossing angle was a result of minor misalignment of the ACC crystals in the delayed branch within the Bragg angle bandwidth.
For speckle measurements that will be presented later, the angles of the asymmetric channel-cut crystals were optimized such that, at $T_0$, with the prism removed, there were no noticeable interference fringes.

We note in addition that even though very distinguished single shot interference patterns were observed at $T_0$, the interference fringe vanishes with multi-pulse average.
This indicates the expected absence of phase stability between the two branches.
According to equation~(\ref{equation:general delay formula}), to achieve phase stability at 9.83~keV, the positioning jitter of the air-bearing stage needs to be much smaller than 0.22~nm (This is the crystal translational motion that corresponds to a $\pi$ phase shift.) and is far smaller than the actual 20~nm positional jitter of the air bearing linear stage.

\section{Performance Evaluation}\label{analysis}
In this section we present analysis of the measured photon throughput and the relative pointing stability between the two branches.
In addition, we investigate the mutual coherence between the two foci in detail by comparing small angle speckle patterns generated from two branches.

We first measured the diffraction and transmission efficiency of $G1$ by imaging the diffraction orders at the sample plane with other components removed.
For both the 0$^\text{th}$ order and $\pm1$ orders of diffraction, the average beam profile of 3000 pulses were used.
Then integrated average intensities are normalized by the upstream beamline $I_0$\cite{Feng2011}.
For $G1$ the estimated efficiencies for -1 ,0, +1 orders at the optimum angle of 72$^\circ$ are determined to be $20.0(2)\%$, $11.0(1)\%$ and $19.0(2)\%$, where the transmission of the 100~$\mu m$ diamond substrate of $79\%$ is included.
For a rectangular-cross-section phase grating with 0.5 duty cycle, neglecting absorption, according to the measured intensity ratio between $\pm1$ and 0$^\text{th}$ order, the phase modulation is about $0.7\pi$.
This leads to an ideal combined efficiency of $26\%$ for the $\pm1$ orders.
The deviation from the ideal efficiency value could be caused by imperfection in the cross section shape of the high aspect ratio structures.

The total throughput of the crystal optics on the SD table was measured with a laser power meter up and down stream of the enclosure, normalized by the upstream $I_0$. 
After accounting for the absorption of vacuum windows, x-ray diagnostic targets, air and helium on along the light path, the total throughput of the two delay lines was estimated to be 38(6)\%.
With the same power meter recording the average pulse energy downstream of $G2$, factoring in the grating efficiency of $\sim$20\% for the first diffraction order, the combined energy efficiency of the split-delay setup was derived to be $1.5(3)\%$. Details of the analysis of the measurement are shown in Appendix~\ref{appendix:grating efficiency} and \ref{appendix:total_energy_efficiency}.

\begin{table}[hbtp]
    \centering
    \begin{tabular}{|l|l|l|l|l|}
    \hline
         & CCs  & Gratings & Others &Total \\ \hline
     Measured  & 42(8)\%   & 8.0(2)\%     & 45(14)\%   &  1.5(3)\%  \\ \hline
     Ideal 1   & 41.5\% & 32\%  & 100\%  & 13.3\% \\ \hline
     Ideal 2   & 81\%   & 32\%  & 100\%  &  25.9\% \\ \hline
    \end{tabular}
    \caption{Transmission efficiency of different components of the setup. The pure energy efficiency of the channel-cut crystals can not be measured directly. The values are derived from a model combined with the measured data. The column "CCs" refers to the transmission efficiency of all channel-cut crystals. The column "Others" refers to the transmission efficiency of everything except the channel-cuts and the two gratings. The "Ideal 1" refers to the energy efficiency with a perfect Gaussian pulse with 0.589~eV bandwidth and "Ideal 2" refers to the energy efficiency of a perfect Gaussian pulse with 0.1~eV bandwidth. }
    \label{table:energy_efficiency}
\end{table}

The comparison between the measured and theoretically optimal throughput is summarized in table \ref{table:energy_efficiency}.
While the measured throughput here is significantly lower than the $21\%$ (9\% and 12\% respectively for the delayed and fixed branch) reported in \cite{li2020design}, within in measurement uncertainty, the reduction can be fully accounted for considering the broader incoming beam bandwidth, actual diamond grating performance parameters, and absorption from beam path in air, x-ray windows, and x-ray diagnostics.
This strongly supports the feasibility of approaching theoretical performance by further improving grating fabrication as well as eliminating air paths and windows.
More details about the calculation can be find in Appendix~\ref{appendix:total_energy_efficiency}.

\begin{figure}[bht!]
    \centering
    \centering
     \includegraphics[width=0.48\textwidth, keepaspectratio]{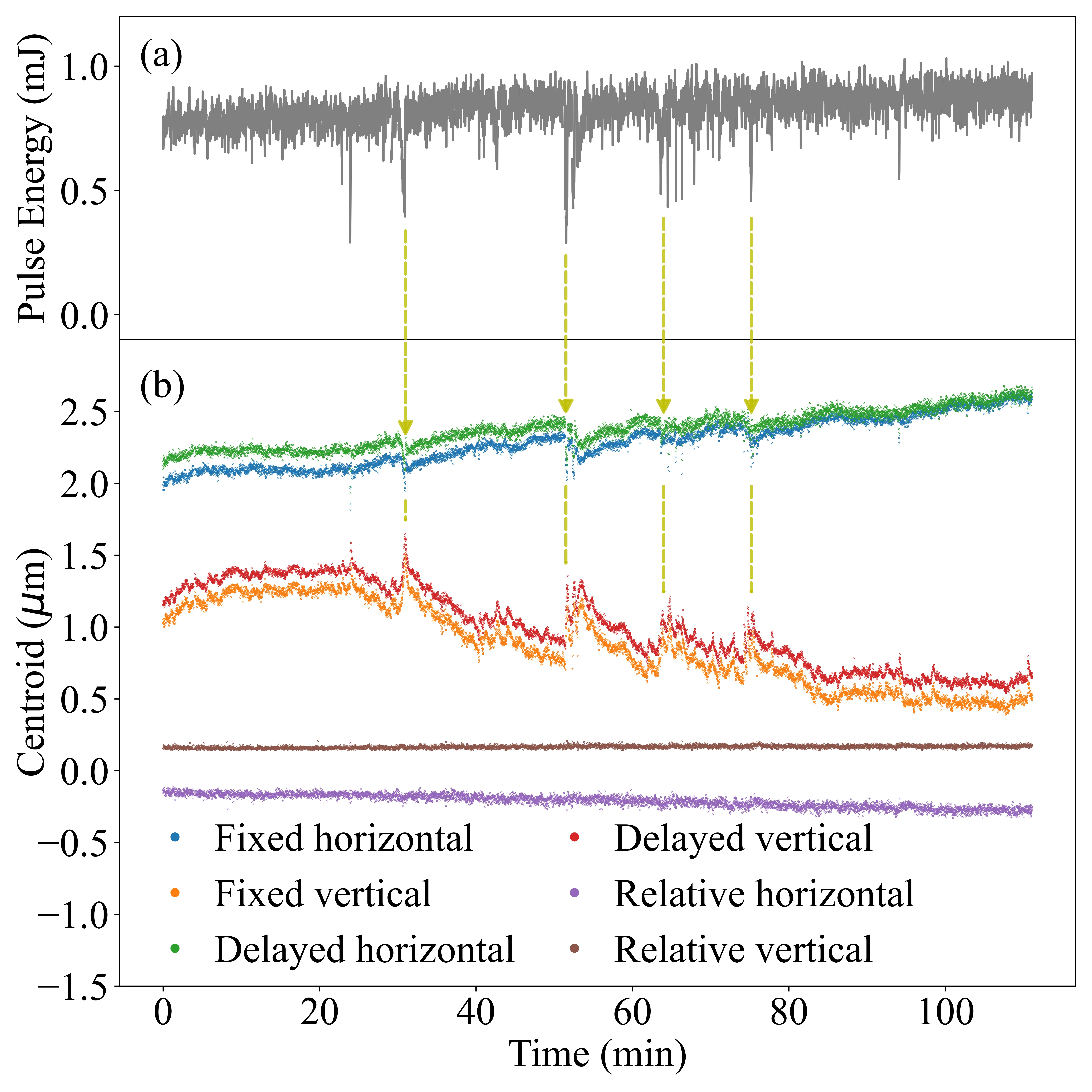}
    \caption{(a) Pulse energy measurement with the upstream gas monitor showing beam delivery interruptions (denoted with arrows) during the long term beam stability measurement. (b) Single branch beam positions and their relative positions over two hours. Each data point reflects the average beam position over 1 second. The yellow dashed arrow and line across (a) and (b) highlights the coincidence between the beam drop and sudden change of the beam positions.}
    \label{fig:profiles}
\end{figure}

Next we discuss the relative stability of the two branches by analyzing the beam positions of the two foci measured in the sample/focal plane.
The two output beams after $G2$ were focused by the CRL with a focal length of $\sim$1~m.
The foci of the two relevant diffraction orders were intentionally separated using the glassy carbon prism between $CC1$ and $CC6$.
The shot-to-shot beam profiles on the sample/focal plane were recorded simultaneously on the scintillator screen.
Shown in Supplement
are examples from a series of consecutive shots, where we observed common fluctuations in both beam position and profile shared between the two branches.
A 2D Gaussian fit was applied to extract the focus centroids for the delayed and fixed branch respectively.
Individual and relative position fluctuation statistics are listed in table~\ref{table:shot-to-shot stability}.
The upstream source and beamline optics instability accounted for majority of the fluctuations, which was reflected in the common movement of both branches.
On the other hand, the system showed a high degree of tolerance to upstream instabilities: the relative position jitter between the two pulses was much smaller than each individual branch.
This can be attributed to the use of amplitude splitting.
The common beam motion is more apparent in long-term focus stability measurements.
As shown in figure~\ref{fig:profiles}, during a 2-hour time span, large amplitude excursions in the single branch positions displayed in (b) coincided largely with the FEL beam delivery interruptions shown in (a).
This relates to the upstream mono thermal instability. 
However, the relative position between the two branches remains robust, with a total drift of $\sim$10~nm and $\sim$130~nm in vertical and horizontal respectively over this two-hour period. 
\begin{table}[hbtp]
    \centering
    \begin{tabular}{|l|l|l|l|}
    \hline
    Centroid (RMS) & Delayed  & Fixed &  Relative \\ \hline
     Horizontal (nm)  & 214 (4)       &   190 (3)  & 50 (5)     \\ \hline
     Vertical (nm) & 146 (4) & 184 (4) &61 (6)\\ \hline
    \end{tabular}
    \caption{Beam shot-to-shot centroid motion.}
    \label{table:shot-to-shot stability}
\end{table}

\begin{figure}[bht!]
    \centering
    \includegraphics[width=0.45\textwidth, keepaspectratio]{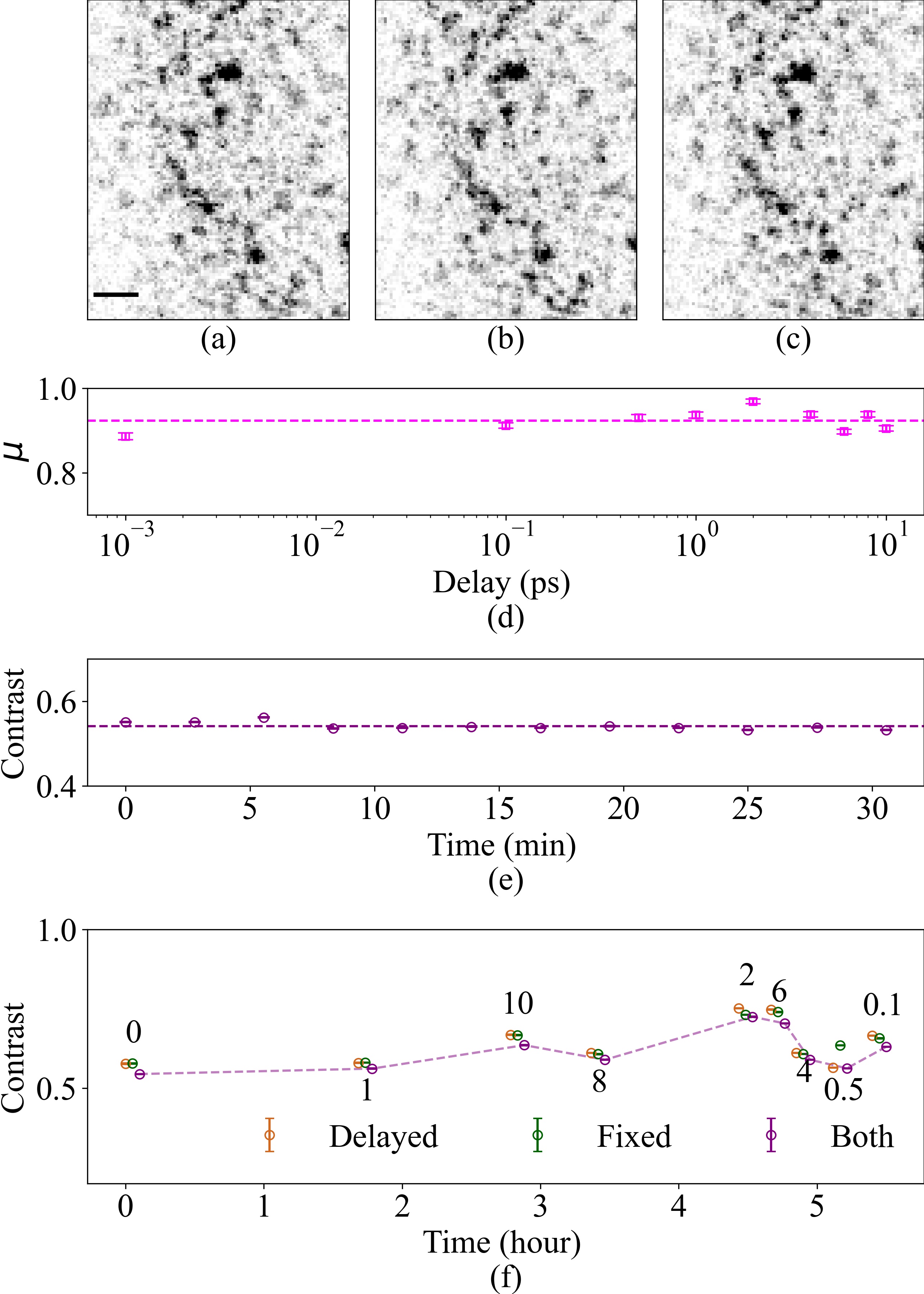}
    \caption{Averaged speckles from delayed (a), fixed (b) and both (c) branches from 50 and 25 frames from single and both branches respectively. The scale bar in (a) is shared by all 3 plots and corresponds to 0.000665 {\AA}$^{-1}$. (d) Measured effective overlap $\mu$ right after we optimized the spatial overlap for each delay. (e) Effective overlap over a time span of 30 minutes. (f) Contrast level of the small angle coherent scattering from delayed, fixed, and both branches as a function of the measurement time. The corresponding delay times are written above/below the markers in the unit of $ps$.}
    \label{fig:Speckle Analysis}
\end{figure}

We finally quantitatively investigate the mutual coherence between the two branches.
The sample-plane scintillator screen is useful for achieving the beam central position overlap of the two beams.
However, it cannot resolve detailed transverse profile of the focused beams due to limited spatial resolution ($\sim1~\mu$m).
Moreover, the beam profiles as well as the spectral content of the pulse pairs fluctuates from pulse to pulse following the input beam variations, potentially impacting the signal quality of an XPCS measurement. 
Therefore, we directly investigate the degree of transverse coherence via analyzing the small angle coherent scattering from a static silica powder sample~\cite{Gutt2012}.
This measurement was performed at 9.5~keV.
In order to decouple the impact of the split-delay optics from upstream fluctuations and optics imperfections, we initially limited the input beam size with a 50~$\mu m$ square aperture.
The ePix100 detector recorded speckle patterns from either both or one of the two branches~\cite{sikorski2016application}.
Visually high contrast and notably similar average speckle patterns were observed in all 3 scenarios as displayed in figure~\ref{fig:Speckle Analysis} (a-c).
To quantify the similarity between the two pulses in the context of XPCS measurements, we evaluate the effective overlap $\mu$, related to visibility degradation in the absence of sample dynamics. It is defined as
\begin{equation}
    \beta = r^2\beta_1+(1-r)^2\beta_2+ 2\mu r(1-r)\mathrm{min}(\beta_1,\beta_2),
\label{eq:effective_overlap}
\end{equation}
where $r$ is the intensity branching ratio of two branches with $r \equiv i_1/(i_1+i_2)$.
The subscript 1,2 of the intensity $i$ and visibility $\beta$ denotes the delayed branch and the fixed branch respectively. 
The angle of $CC6$ was detuned to achieve an equal intensity splitting ($r\approx 0.5$) between the two branches.
Several delay points spanning over the $\sim$10~ps time delay range of the system were selected for scattering measurements. 
For each delay point, a two-step visibility analysis was performed to get the visibility in the 3 conditions corresponding to $r = 0, 0.5, 1$.
First, we used the droplet based 'greedy guess' algorithm to locate photon positions in each speckle pattern~\cite{sun2020accurate}. 
Then, from a large number of frames, a maximum likelihood estimator was applied to find the mode number that optimizes the likelihood of the negative binomial distribution from our photon statistics measurements, \emph{i.e.}, the probabilities of 1, 2, and 3 photons per pixel within the count rate range from 0.01 - 0.1 photon per pixel~\cite{hruszkewycz2012high,roseker2018towards}. 

The calculated effective overlap is plotted in figure~\ref{fig:Speckle Analysis} (d) for the selected delay points.
It is consistently above 90\%.
We note that before focusing the two output beams had a 22~$\mu m$ horizontal relative motion when translating $CC2$ and $CC3$ together over a 10~mm scan, potentially arising from a $\sim 0.02^\circ$ asymmetry angle mismatch between $CC2$ and $CC3$.
This led to a $\sim$200~nm relative horizontal motion between the two beams at focus (See Appendix~\ref{appendix:AsymmetryAngleMismatch}
 for details). 
We compensated this beam offset by translating $CC5$ at each time delay as an optimization procedure to achieve optimal spatial overlap. 
At each fixed time delay, the summed-speckle contrast experienced less than 3\% change during our half an hour measurement as plotted in figure~\ref{fig:Speckle Analysis} (e). 
However, as shown in figure~\ref{fig:Speckle Analysis} (f), single-branch contrast values showed non-negligible variations across different time delays measured during a period of several hours.
This implies that, even though the two branches can maintain a sufficient level of relative stability, the upstream beam condition variation can impact individual beam's transverse properties, e.g., the upstream beam trajectory/position drift may change the transverse portion of the beam that illuminates the slit and the lens.
This poses challenges in the data interpretation/normalization and can potentially lead to systematic errors in an actual XPCS measurement, in which intrinsic dynamics is also revealed through contrast changes. 

\begin{figure}[bht!]
    \centering
    \includegraphics[width=0.45\textwidth, keepaspectratio]{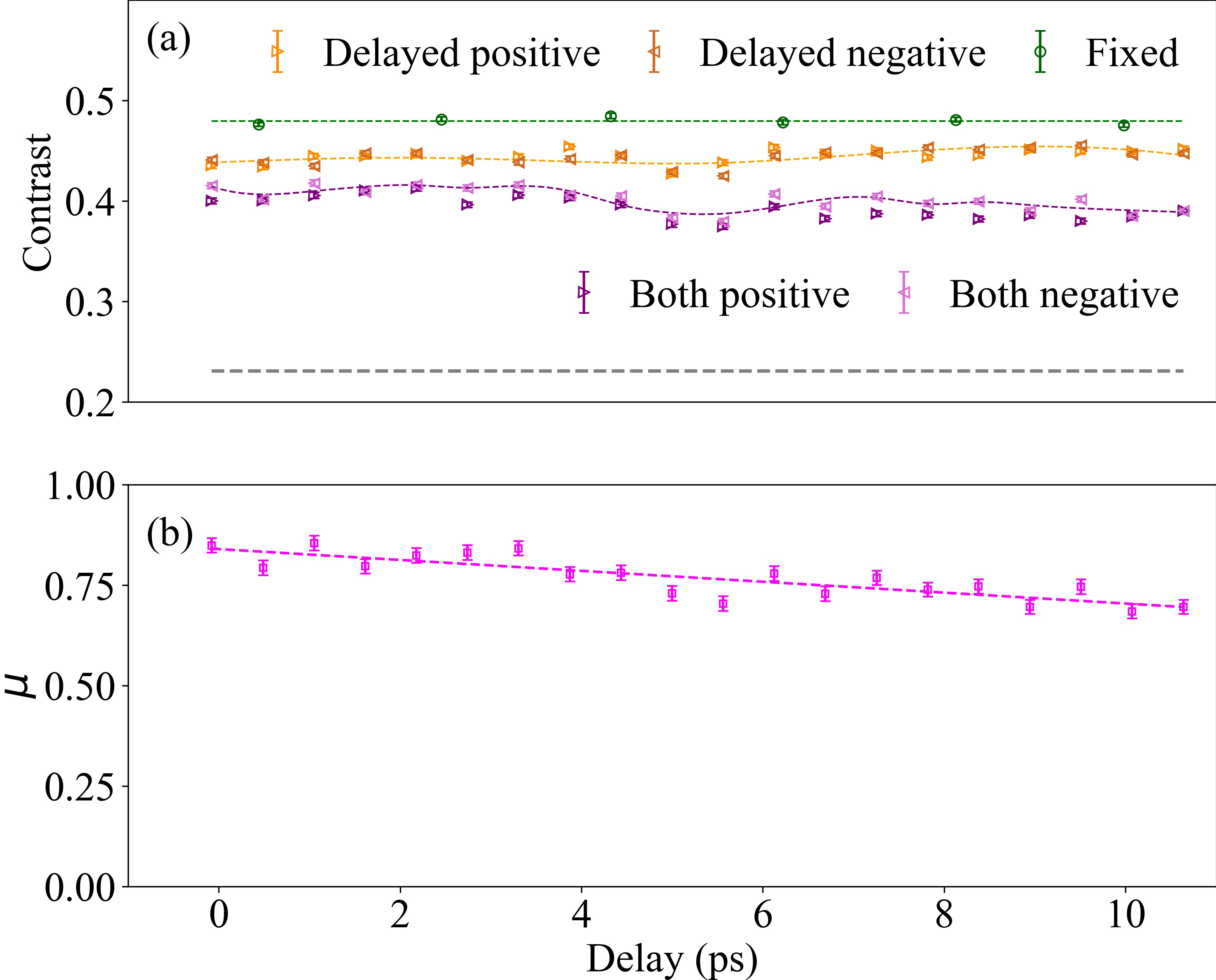}
    \caption{(a) Contrast of the small angle coherent scattering from delayed, fixed and both branches in the continuous delay scan mode with the calculated effective overlap displayed in (b). The gray dashed line in (a) is the projected contrast curve for $\mu$ = 0, calculated with the averaged delayed and fixed branch contrast.}
    \label{fig:delay scan}
\end{figure}

To overcome these types of drifts over the time scale of minutes and hours, a better scheme of measurement is to repeat time delays faster than the time scale of these drifts~\cite{glownia2019pump, sun2019compact}. 
We thus performed fly-scan test with a speed of 0.3~mm/s or 0.28~ps/s.
The contrast curves from the 3 conditions, extracted from scattering patterns which are grouped based on delay times, are plotted in figure~\ref{fig:delay scan} (a).
A different slit setting was used to mitigate effects of beam relative offsets due to the gap mismatch between $CC2$ and $CC3$: the upstream slits were wide open, the slits right upstream of the CRLs were closed down to 150~$\mu m$ so as to always illuminate the same area on the lens.
The contrast values in this configuration are noticeably lower than those presented figure~\ref{fig:Speckle Analysis}.
This could be attributed to imperfections of upstream optics such as the known asymmetry angle in the beamline monochromator diamond crystal \cite{zhu2014performance}.
On the other hand, individual branches are more similar across different delays, as we see significantly less contrast variations from each branch as a function of delay time.
A bi-directionality was also observed, manifested in the small difference in contrast levels in the positive (delay increase) and negative (delay decrease) scan directions, likely due to cable tension.
Nevertheless, as displayed in figure~\ref{fig:delay scan} (b), the overall effective overlap maintains at a high level, showing negligible changes in $\mu$ within the first 4~ps.

\section{Conclusion}
In summary, we have experimentally implemented the new x-ray split-delay line using a grating-based amplitude-splitting all-channel cut design.
The system is able to generate femtosecond x-ray pulse pairs with significantly higher mutual coherence compared to previously realized systems.
This is manifested in both the high contrast interference fringes and the high contrast two-pulse coherent small angle scattering patterns.
We have also demonstrated the expected high relative stability between the two branches, which is well preserved in spite of the incoming x-ray beam pointing drift.
The overlap stability during continuous delay scans, enabled by the channel-cut crystal pair translation with a single air bearing stage, 
allows fast and accurate delay-time repetition, which is essential for robust high-sensitivity time domain measurements.
Being able to maintain the overlap between micron sized x-ray beams makes it possible to perform x-ray pump x-ray probe experiments with higher intensity x-ray excitation, e.g. enabling the generation and diagnosis of warm dense matters. 
Albeit covering only a relatively small time window of $\sim$10 picoseconds, systematic exploration of ultrafast dynamics in disordered matters on the picosecond scale with sub-100~fs time resolution through speckle visibility spectroscopy also becomes feasible.

We also note that the amplitude-splitting concept can be adopted to most existing split-delay optical systems by the introduction of grating beam splitters up and down stream of the delay lines.
While one would anticipate a reduction of available flux at the sample, this is more than compensated for by the improvement in mutual coherence which will increase the signal to noise and signal to background ratio for most cases significantly.

The all-channel-cut system can also, in principle, be expanded to cover larger delay time ranges by adopting artificial channel-cut crystals with longer reflecting surfaces.
This will ease the crystal manufacturing requirements and potentially yield higher surface quality as well as more accurate crystal asymmetry angle control.
With the introduction of moderate cooling to the crystals, we anticipate this as a viable path towards deploying the split-pulse XPCS methodology as a robust way for studying disorder and fluctuations at the atomic scale at the upcoming high repetition rate sources.

\begin{acknowledgments}
Use of the LCLS at SLAC National Accelerator Laboratory, is supported by the U.S. Department of Energy, Office of Science, Office of Basic Energy Sciences under Contract No. DE-AC02-76SF00515.

\end{acknowledgments}

\appendix

\section{Grating Diffraction Efficiency}\label{appendix:grating efficiency}

\begin{figure}[bht!]
    \centering
    \includegraphics[width=0.4\textwidth, keepaspectratio]{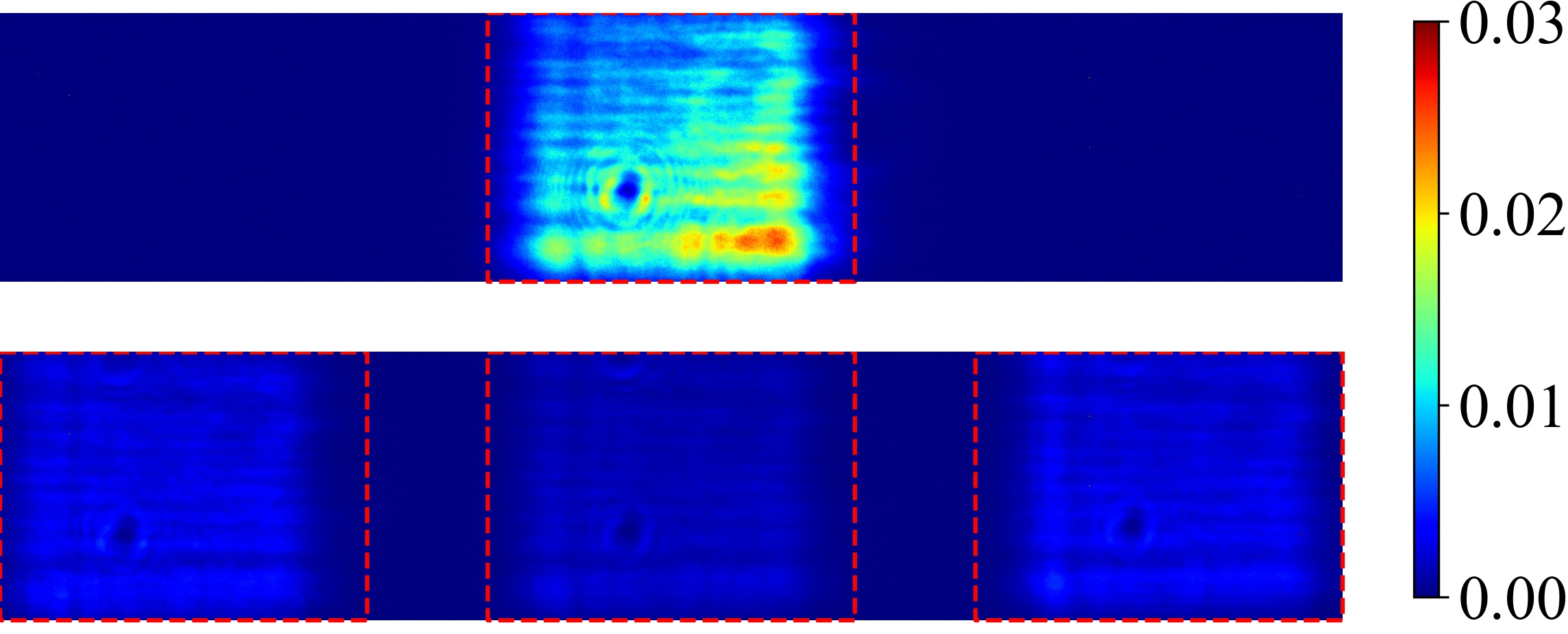}
    \caption{(Top) Direct beam intensity profile on the beam profile monitor. We calculate the total intensity within the red boxes.  The size of the box is $700\times 512$ pixels. (Bottom) The -1, 0 and 1 order of diffraction (left, middle and right) from the first grating. We calculate the total intensity within the red boxes. The size of the red boxes is $700\times 512$ pixels. }
    \label{fig:after_g1}
\end{figure}
Scintillator screen images for the direct beam (up) and diffracted beams from $G1$ (down) are shown in figure~\ref{fig:after_g1}.
We calculate the pulse intensity in the red boxes.
Then the intensities are normalized with the incident pulse beam energy measured with upstream intensity meter. The ratios between the normalized intensities are defined as transmission efficiencies.
Detailed calculation can be found in the code base \cite{code}.

\section{Total Energy Transmission Efficiency}\label{appendix:total_energy_efficiency}
\subsection{Simulation}
Detailed calculation of the energy transmission efficiency can be found in the code base \cite{code}.
Here we explain the procedure of the calculation.
First, assume the incident pulse is a perfect 3D Gaussian pulse with a spectral bandwidth of 0.589~eV, the full-width-half-maximum of the intensity reflectivity curve of diamond (111) Bragg reflection at 9.83~keV.
The corresponding bandwidth limited pulse duration is 1.55~fs.
The spatial size is set to be  150~$\mu$m and the central energy is set to be 9.5~keV.
With the simulation setup, we calculate the output electric field from the device and obtain the ideal energy efficiency without any air, Kapton tape, or other kinds of absorption.
Especially, during this process, the grating shapes are assumed to be perfect. Therefore, the grating efficiency value 40\% is used here.
At this stage, the energy efficiency of the delayed branch is 5.7\% while the energy efficiency of the fixed branch is 7.6\%.
Then we calculate the energy loss due to the imperfection of the grating shape, diamond substrate of the gratings, air, Kapton tapes, and diamond scintillator upstream and downstream of the SD table.
The energy efficiency of each components is shown in table~\ref{table:energy_efficiency_components}.

\begin{table}[hbtp]
    \centering
    \begin{tabular}{|l|l|l|l|}
    \hline
    & l($\mu m$) & $\eta(\mu m^{-1})$ & Efficiency \\ \hline
     Grating   & N/A      &   N/A                   &  63 \%  \\ \hline
     Substrate & 324(40)    &  $7.3\times10^{-4} $    &  79(2) \%  \\ \hline
     Air       & $6.2(1)\times 10^5$ & $6.2\times10^{-5} $  &  68(4) \%\\ \hline
    Kapton D    & 455(10)    &   $4.43\times 10^{-4}$  &  81.7(4)  \% \\ \hline
    Kapton DF   & 280(10)    &  $4.43\times 10^{-4}$   &  88.3(4)\% \\ \hline
   Scintillator&  60(2)   &    $7.3\times10^{-4}$   &  95.6(2)\% \\ \hline
    \end{tabular}
    \caption{Energy efficiency table for each components. 
    }
    \label{table:energy_efficiency_components}
\end{table}

In the table~\ref{table:energy_efficiency_components}, the "Grating" refers to the deviation from a perfect $\pi$ phase grating with 0.5 duty cycle. 
Ideally, a $\pi$ phase grating has energy efficiency of $40\%$ in $\pm 1$ orders of diffraction.
In this experiment, the measured value is $20\%$.
Therefore, considering the absorption of the grating substrate, the energy efficiency compared with the ideal case is $0.2 / 0.79 / 0.4 \approx 63\% $ where 0.79 is the energy efficiency of the grating substrate at $72^\circ$.
The "Substrate" refers to the effective substrate thickness including the tilting angle of $72^\circ$.
The "Air" refers to the air gaps outside the helium cover and the effective gap inside the helium cover.
The total path length inside the helium cover is about 120cm.
Assume that the percentage of helium is 90\%.
Therefore, the estimated effective air path inside the helium cover is 12~cm. The "Kapton D" refers to Kapton films for the delayed branch while "Kapton DF" refers to the fixed branch.
"Scintillator" refers to the two diamond scintillators upstream and downstream of the SD table for beam profile monitoring: each has a thickness of 30~$\mu m$.
$\eta$ refers to the absorption length and $l$ is the length of the material.
The transmission is calculated with the formula $\exp{(-\eta l)}$.

Note that in the table the uncertainty of the lengths is given with estimation rather than actual measurement.
The corresponding uncertainty in the efficiency is assumed to be half of the change of the efficiency over the uncertainty region of the length.
Therefore, with this table, the theoretical energy efficiency is calculated for the delayed and fixed branches respectively. 
For the delayed branch, 
\begin{align*}
        E_{Delayed} &= 5.7\% \times (63\% \times 79\%)^2 \times 82\% \times 96\% \times 68\% \\
        &\approx 0.76\%.
\end{align*}
For the fixed branch
\begin{align*}
        E_{Fixed} &= 7.6\% \times (63\% \times 79\%)^2 \times 88\% \times 96\% \times 68\% \\
        &\approx 1.08\%.
\end{align*}
Therefore, the theoretical prediction of the total energy efficiency is $0.76\% + 1.08\% = 1.8\%$.
The uncertainty of the total energy efficiency is obtained with the standard formula for uncertainty propagation.

\subsection{Measurement}
Right after the second grating, the directly measured energy efficiency is $5.99(2)\%$.
The 0.02\% error is statistical.
The relative systemic error here can be up to 10\%.
The energy efficiency is estimated to be $6.0(6)\%$. 
This $6.0(6)\%$ throughput was obtained when the slit after the second grating was fully open.
Assume that the angle of the second grating is $72(2)^\circ$ and $\pm 1$ orders of diffraction have a transmission of $20(2)\%$. The total energy transmission efficiency parallel to the incident pulse and available for XPCS measurement is estimated to be $1.8(3)\%$.
The details of the calculation can be found in the code base \cite{code}.

\subsection{Energy Efficiency of Channel-cuts}
It is challenging to directly measure the energy efficiency of the channel-cut crystals alone in current setting.
Therefore, the channel-cut efficiency shown in the body text is obtained through estimation based on a model, which will be explained in detail below.
In the experiment, compared with the incident pulse energy, the energy measured right after the SD table is $11(2)\%$.
The theoretical throughput right upstream the SD table is calculated to be 70(2)\%, including absorption from Kapton films, beam paths through air, and the grating substrate.
Compared with this value and with the measured diffraction efficiency of the $\pm 1$ orders of diffraction of the first grating, the energy efficiency of the SD table is estimated to be 32(5)\%.
Between the two measurement positions, the absorption materials and the corresponding energy transmission efficiency is listed in table~\ref{table:energy_efficiency_components_SD_table}.

\begin{table}[hbtp]
    \centering
    \begin{tabular}{|l|l|l|l|}
    \hline
    & l($\mu m$) & $\eta(\mu m^{-1})$ & Efficiency \\ \hline
     Air       & $2.20(5)\times 10^5$ & $6.2\times10^{-5} $  &  87(3) \%\\ \hline
    Kapton Mean    & 200(100)      &   $4.43\times 10^{-4}$  &  90(8)  \% \\ \hline
   Scintillator&  60(2)      &    $7.3\times10^{-4}$   &  95.6(2)\% \\ \hline
    \end{tabular}
    \caption{Energy efficiency table for each components. 
    }
    \label{table:energy_efficiency_components_SD_table}
\end{table}
Therefore, the energy efficiency of the channel-cut crystals is calculated to be 42(8)\%.
The energy transmission efficiency of all other components can be represented as
\begin{equation}
    f_{Measured} = 1.5\% / 42\% / 20\% / 40\% \approx 50\%
\end{equation}
Detailed calculation of the uncertainty can be found in the code base \cite{code}.

\section{Prism Steering Angle}\label{apopendix:prism steering}
\begin{figure}[bht!]
    \centering
    \includegraphics[width=0.45\textwidth, keepaspectratio]{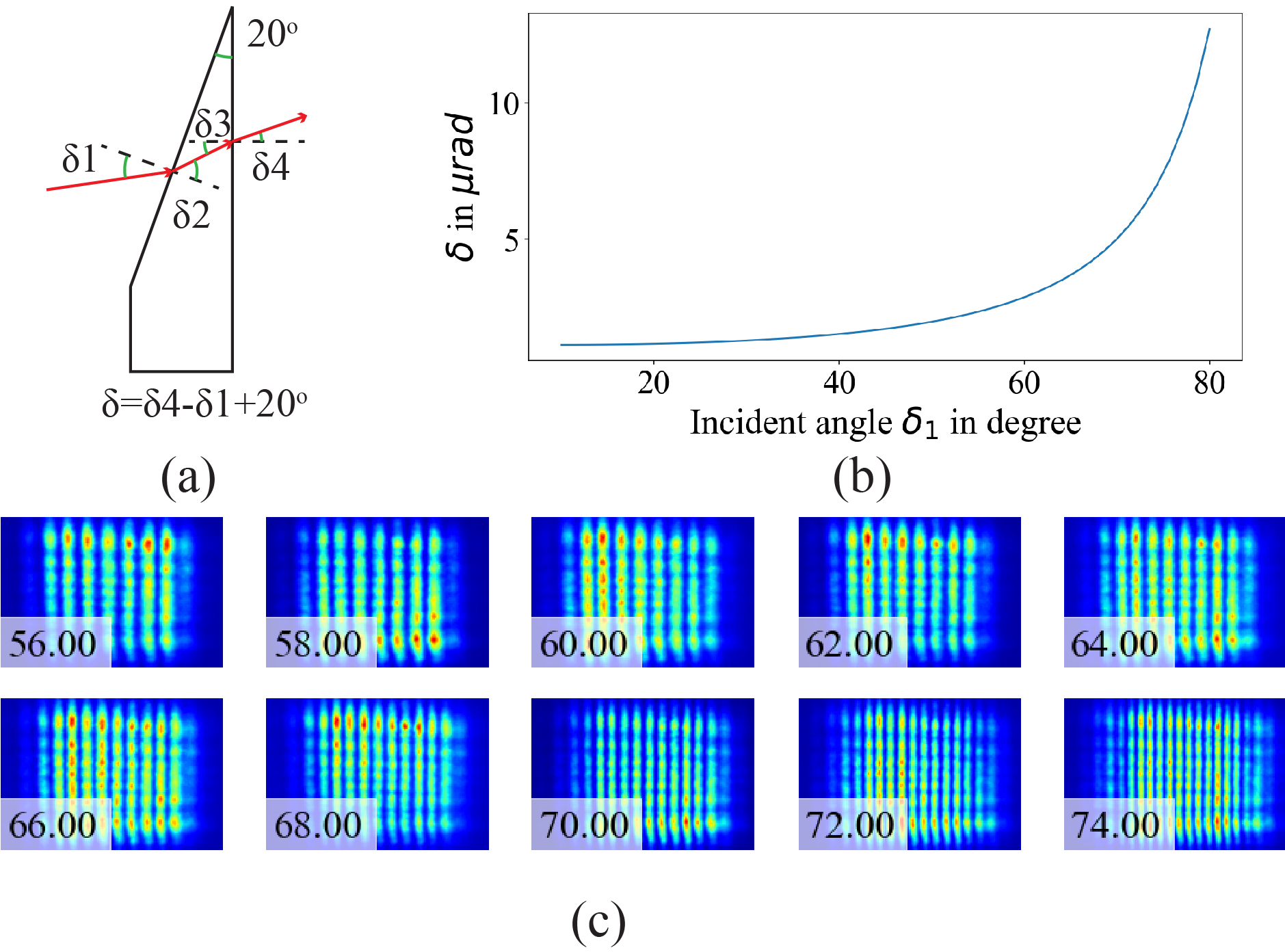}
    \caption{(a) Structure of the glassy carbon prism. (b) Steering angle of the prism for 9.83~keV photons for different incident angle. (c) Interference patterns measured with different prism angles. The numbers on the lower left corner for each image is the estimated incident angle of the prism.}
    \label{fig:prism_structure}
\end{figure}
 
The geometric shape of the glassy carbon prism is shown in figure~\ref{fig:prism_structure}(a). The prism angle is $20^\circ$.
The density of the glassy carbon is $1.5g/cm^3$ according to the manufacturer. 
The steering angle of the prism for 9.83 keV photons is shown in figure~\ref{fig:prism_structure}(b), where the angles are defined in figure~\ref{fig:prism_structure}(a).
In this experiment, the incident angle is estimated to be between $69^\circ$ and $74^\circ$. The corresponding steering angle range 4.7 to 6.8~$\mu rad$. 
In figure~\ref{fig:prism_structure}(c) we show the interference patterns for several different prism angles.

\section{Locating $T_0$}\label{appendix:t0 configuration}
For Si (220) Bragg reflection, $5^\circ$ asymmetry angle and 9.5keV photon energy, during the delay scan, according to the general equation ~(\ref{equation:general delay formula}) the delay time change is related to the air-bearing stage displacement $\Delta d$ through the formula
\begin{equation}
    \Delta(t)\ (ps) =  0.940\ \Delta (d)\ (mm)
    \label{eq:delay formula}
\end{equation} 

According to equation~(\ref{eq:delay formula}), the relative delay time is sensitive to the delay-scan stage position.
As measured with the visibility curve in figure 2(a), the coherence time is $\sim$11~fs. 
Therefore, the interference fringe is prominent within a 12~$\mu m$ position range.
With the installation accuracy of various components in the system, the remaining uncertainty of $T_0$ corresponds to a $\Delta d$ range of 1mm.
We can locate $T_0$ rather quickly by moving the air bearing stage near the nominal position.
The $T_0$ configuration also depends on the incident photon energy.
For example, changing from 9.83 keV to 9.5 keV, the position of the air-bearing stage changes about 4~mm in the $T_0$ configuration.

\section{Python Simulation}\label{appendix:python_simulator}
All simulation results mentioned in the body text are obtained through a python simulator contained in the code base \cite{code}.
In this simulation, the incident pulse is assumed to be a perfect 3D Gaussian pulse. Then the diffraction and propagation phenomena of the pulse are simulated in the wave-vector space for each monochromatic component.
The Bragg reflection from the crystals is calculated with 2-beam dynamical diffraction theory. For gratings, a perfect $\pi$ phase shift grating is assumed.
For the implementation, a GPU accelerator is used to accelerate the calculation. 
To make the simulation more realistic, a CAD model of the system is implemented with \emph{Solid Edge 2019}.
We have assembled the system according to this model.
The positions of the crystals in the model are a good approximation of the actual positions in experiment.
We translate the positions in the CAD model to NumPy arrays and use that for the simulation.

\section{Impact of Alignment Errors}\label{appendix:InfluenceofAlignmentError}
The manufacture of the channel-cut crystals and our alignment are not perfect. In this section, we discuss their influence on the performance on this setup.
The discussion is based on the python simulator mentioned above.

\subsection{Pure Alignment Error}
When ideally aligned, this device has no overall angular dispersion. 
However, when the alignment is not perfect, overall angular dispersion appears.
Assume that the delayed branch is detuned by 5.2~$\mu$rad.
Then during a 10~mm delay scan, the relative position of the two branches change 20~nm horizontally before focusing. 
If additionally, there is also a misalignment out of the diffraction plane (i.e. the $x-z$ plane), then during a 10~mm delay scan, the relative position will change in both horizontal and vertical directions.
For example, if $CC2$ is rotated by $0.01^\circ$ around its long edge, then during a 10~mm delay scan, the relative position between two branches will change 24~nm horizontally and 743~nm vertically.

\subsection{Asymmetry Angle Mismatch}\label{appendix:AsymmetryAngleMismatch}
\begin{figure}[bht!]
    \centering
    \includegraphics[width=0.45\textwidth, keepaspectratio]{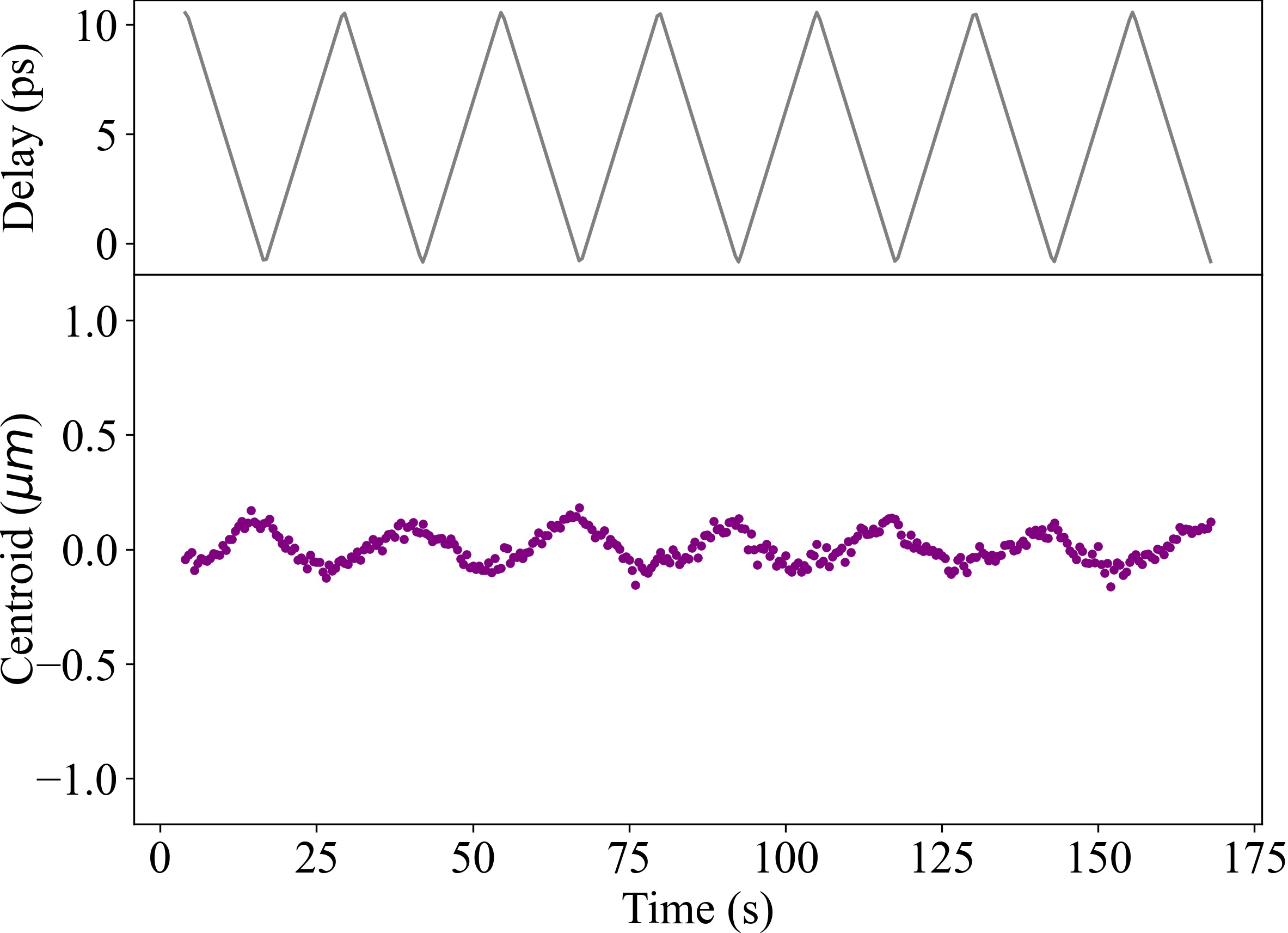}
    \caption{Horizontal motion of the focused delay branch beam during the continuous delay scan.}
    \label{fig:delay_scan_wobble}
\end{figure}

Previously, we assume asymmetry angles are exactly $5.00^\circ$ for all asymmetric channel-cut crystals.
Here We show the impact of a small mismatch between the asymmetry angles within a CC pair.
Assume that there is a $0.02^\circ$ mismatch between $CC2$ and $CC3$, the 10~mm delay scan induces a 22~$\mu m$ horizontal change of the relative position between the two pulses.
If $CC2$ also has a $0.01^0$ misalignment around it long axis out off the diffraction plane, then the 10~mm delay scan will lead to a $22~\mu$m horizontal change and a $744$~nm vertical change of the relative position.
This can explain the cyclic relative motion during our delay scan.
In a $\sim$ 10~ps delay scan, we observed a cyclic horizontal motion of the unfocused delayed branch beam. 
With the focusing optics, the position error is demagnified to be $\sim$200~nm peak to peak beam wobble, in agreement with our measurement shown in figure~\ref{fig:delay_scan_wobble}. 

\subsection{Grating Misalignment}
The misalignment of the orientation of the first grating has very limited influence of the properties of the setup.
Assume that $G1$ is misaligned by $1^\circ$ in the $x-y$ plane, the position of the delayed branch focus will changes 27~nm horizontally and 12~nm vertically during a full-range delay scan. 
Therefore, for the alignment of $G1$, adjustment based on unfocused beams on scintillator screens should be good enough.

\section{Delay time range}\label{appendix:DelayTimeRange}
The delay time range is determined by the crystal size, and according equation~(\ref{equation:general delay formula}) influenced by the asymmetry angle and Bragg angle. 

For this specific device, the following delay range is estimated with the python simulator with actual crystal parameters. In this table, positive delay time means that the fixed branch arrives before the delayed branch.

\begin{table}[htp]
\centering
\begin{tabular}{|l|l|}
\hline
Photon Energy & Delay range      \\ \hline
7~keV                  & -2~ps to 18~ps     \\ \hline
9.5~keV                  & -0.5~ps to 13.5~ps \\ \hline
12~keV                   & -9.8~ps to 0.5~ps  \\ \hline
\end{tabular}
\caption{Estimated delay range of this device for different photon energies.}
\label{table:delayRangeEstimation}
\end{table}

\bibliography{apssamp}

\begin{thebibliography}{36}%
\makeatletter
\providecommand \@ifxundefined [1]{%
 \@ifx{#1\undefined}
}%
\providecommand \@ifnum [1]{%
 \ifnum #1\expandafter \@firstoftwo
 \else \expandafter \@secondoftwo
 \fi
}%
\providecommand \@ifx [1]{%
 \ifx #1\expandafter \@firstoftwo
 \else \expandafter \@secondoftwo
 \fi
}%
\providecommand \natexlab [1]{#1}%
\providecommand \enquote  [1]{``#1''}%
\providecommand \bibnamefont  [1]{#1}%
\providecommand \bibfnamefont [1]{#1}%
\providecommand \citenamefont [1]{#1}%
\providecommand \href@noop [0]{\@secondoftwo}%
\providecommand \href [0]{\begingroup \@sanitize@url \@href}%
\providecommand \@href[1]{\@@startlink{#1}\@@href}%
\providecommand \@@href[1]{\endgroup#1\@@endlink}%
\providecommand \@sanitize@url [0]{\catcode `\\12\catcode `\$12\catcode
  `\&12\catcode `\#12\catcode `\^12\catcode `\_12\catcode `\%12\relax}%
\providecommand \@@startlink[1]{}%
\providecommand \@@endlink[0]{}%
\providecommand \url  [0]{\begingroup\@sanitize@url \@url }%
\providecommand \@url [1]{\endgroup\@href {#1}{\urlprefix }}%
\providecommand \urlprefix  [0]{URL }%
\providecommand \Eprint [0]{\href }%
\providecommand \doibase [0]{https://doi.org/}%
\providecommand \selectlanguage [0]{\@gobble}%
\providecommand \bibinfo  [0]{\@secondoftwo}%
\providecommand \bibfield  [0]{\@secondoftwo}%
\providecommand \translation [1]{[#1]}%
\providecommand \BibitemOpen [0]{}%
\providecommand \bibitemStop [0]{}%
\providecommand \bibitemNoStop [0]{.\EOS\space}%
\providecommand \EOS [0]{\spacefactor3000\relax}%
\providecommand \BibitemShut  [1]{\csname bibitem#1\endcsname}%
\let\auto@bib@innerbib\@empty
\bibitem [{\citenamefont {Seddon}\ \emph {et~al.}(2017)\citenamefont {Seddon},
  \citenamefont {Clarke}, \citenamefont {Dunning}, \citenamefont
  {Masciovecchio}, \citenamefont {Milne}, \citenamefont {Parmigiani},
  \citenamefont {Rugg}, \citenamefont {Spence}, \citenamefont {Thompson},
  \citenamefont {Ueda} \emph {et~al.}}]{seddon2017short}%
  \BibitemOpen
  \bibfield  {author} {\bibinfo {author} {\bibfnamefont {E.}~\bibnamefont
  {Seddon}}, \bibinfo {author} {\bibfnamefont {J.}~\bibnamefont {Clarke}},
  \bibinfo {author} {\bibfnamefont {D.}~\bibnamefont {Dunning}}, \bibinfo
  {author} {\bibfnamefont {C.}~\bibnamefont {Masciovecchio}}, \bibinfo {author}
  {\bibfnamefont {C.}~\bibnamefont {Milne}}, \bibinfo {author} {\bibfnamefont
  {F.}~\bibnamefont {Parmigiani}}, \bibinfo {author} {\bibfnamefont
  {D.}~\bibnamefont {Rugg}}, \bibinfo {author} {\bibfnamefont {J.}~\bibnamefont
  {Spence}}, \bibinfo {author} {\bibfnamefont {N.}~\bibnamefont {Thompson}},
  \bibinfo {author} {\bibfnamefont {K.}~\bibnamefont {Ueda}}, \emph {et~al.},\
  }\bibfield  {title} {\bibinfo {title} {Short-wavelength free-electron laser
  sources and science: a review},\ }\href@noop {} {\bibfield  {journal}
  {\bibinfo  {journal} {Reports on Progress in Physics}\ }\textbf {\bibinfo
  {volume} {80}},\ \bibinfo {pages} {115901} (\bibinfo {year}
  {2017})}\BibitemShut {NoStop}%
\bibitem [{\citenamefont {Yamauchi}\ \emph {et~al.}(2002)\citenamefont
  {Yamauchi}, \citenamefont {Mimura}, \citenamefont {Inagaki},\ and\
  \citenamefont {Mori}}]{yamauchi2002figuring}%
  \BibitemOpen
  \bibfield  {author} {\bibinfo {author} {\bibfnamefont {K.}~\bibnamefont
  {Yamauchi}}, \bibinfo {author} {\bibfnamefont {H.}~\bibnamefont {Mimura}},
  \bibinfo {author} {\bibfnamefont {K.}~\bibnamefont {Inagaki}},\ and\ \bibinfo
  {author} {\bibfnamefont {Y.}~\bibnamefont {Mori}},\ }\bibfield  {title}
  {\bibinfo {title} {Figuring with subnanometer-level accuracy by numerically
  controlled elastic emission machining},\ }\href@noop {} {\bibfield  {journal}
  {\bibinfo  {journal} {Review of scientific instruments}\ }\textbf {\bibinfo
  {volume} {73}},\ \bibinfo {pages} {4028} (\bibinfo {year}
  {2002})}\BibitemShut {NoStop}%
\bibitem [{\citenamefont {Yamauchi}\ \emph {et~al.}(2011)\citenamefont
  {Yamauchi}, \citenamefont {Mimura}, \citenamefont {Kimura}, \citenamefont
  {Yumoto}, \citenamefont {Handa}, \citenamefont {Matsuyama}, \citenamefont
  {Arima}, \citenamefont {Sano}, \citenamefont {Yamamura}, \citenamefont
  {Inagaki} \emph {et~al.}}]{yamauchi2011single}%
  \BibitemOpen
  \bibfield  {author} {\bibinfo {author} {\bibfnamefont {K.}~\bibnamefont
  {Yamauchi}}, \bibinfo {author} {\bibfnamefont {H.}~\bibnamefont {Mimura}},
  \bibinfo {author} {\bibfnamefont {T.}~\bibnamefont {Kimura}}, \bibinfo
  {author} {\bibfnamefont {H.}~\bibnamefont {Yumoto}}, \bibinfo {author}
  {\bibfnamefont {S.}~\bibnamefont {Handa}}, \bibinfo {author} {\bibfnamefont
  {S.}~\bibnamefont {Matsuyama}}, \bibinfo {author} {\bibfnamefont
  {K.}~\bibnamefont {Arima}}, \bibinfo {author} {\bibfnamefont
  {Y.}~\bibnamefont {Sano}}, \bibinfo {author} {\bibfnamefont {K.}~\bibnamefont
  {Yamamura}}, \bibinfo {author} {\bibfnamefont {K.}~\bibnamefont {Inagaki}},
  \emph {et~al.},\ }\bibfield  {title} {\bibinfo {title} {Single-nanometer
  focusing of hard x-rays by kirkpatrick--baez mirrors},\ }\href@noop {}
  {\bibfield  {journal} {\bibinfo  {journal} {Journal of Physics: Condensed
  Matter}\ }\textbf {\bibinfo {volume} {23}},\ \bibinfo {pages} {394206}
  (\bibinfo {year} {2011})}\BibitemShut {NoStop}%
\bibitem [{\citenamefont {Seiboth}\ \emph {et~al.}(2017)\citenamefont
  {Seiboth}, \citenamefont {Schropp}, \citenamefont {Scholz}, \citenamefont
  {Wittwer}, \citenamefont {R{\"o}del}, \citenamefont {W{\"u}nsche},
  \citenamefont {Ullsperger}, \citenamefont {Nolte}, \citenamefont
  {Rahom{\"a}ki}, \citenamefont {Parfeniukas} \emph
  {et~al.}}]{seiboth2017perfect}%
  \BibitemOpen
  \bibfield  {author} {\bibinfo {author} {\bibfnamefont {F.}~\bibnamefont
  {Seiboth}}, \bibinfo {author} {\bibfnamefont {A.}~\bibnamefont {Schropp}},
  \bibinfo {author} {\bibfnamefont {M.}~\bibnamefont {Scholz}}, \bibinfo
  {author} {\bibfnamefont {F.}~\bibnamefont {Wittwer}}, \bibinfo {author}
  {\bibfnamefont {C.}~\bibnamefont {R{\"o}del}}, \bibinfo {author}
  {\bibfnamefont {M.}~\bibnamefont {W{\"u}nsche}}, \bibinfo {author}
  {\bibfnamefont {T.}~\bibnamefont {Ullsperger}}, \bibinfo {author}
  {\bibfnamefont {S.}~\bibnamefont {Nolte}}, \bibinfo {author} {\bibfnamefont
  {J.}~\bibnamefont {Rahom{\"a}ki}}, \bibinfo {author} {\bibfnamefont
  {K.}~\bibnamefont {Parfeniukas}}, \emph {et~al.},\ }\bibfield  {title}
  {\bibinfo {title} {Perfect x-ray focusing via fitting corrective glasses to
  aberrated optics},\ }\href@noop {} {\bibfield  {journal} {\bibinfo  {journal}
  {Nature Communications}\ }\textbf {\bibinfo {volume} {8}},\ \bibinfo {pages}
  {1} (\bibinfo {year} {2017})}\BibitemShut {NoStop}%
\bibitem [{\citenamefont {Mimura}\ \emph {et~al.}(2014)\citenamefont {Mimura},
  \citenamefont {Yumoto}, \citenamefont {Matsuyama}, \citenamefont {Koyama},
  \citenamefont {Tono}, \citenamefont {Inubushi}, \citenamefont {Togashi},
  \citenamefont {Sato}, \citenamefont {Kim}, \citenamefont {Fukui} \emph
  {et~al.}}]{mimura2014generation}%
  \BibitemOpen
  \bibfield  {author} {\bibinfo {author} {\bibfnamefont {H.}~\bibnamefont
  {Mimura}}, \bibinfo {author} {\bibfnamefont {H.}~\bibnamefont {Yumoto}},
  \bibinfo {author} {\bibfnamefont {S.}~\bibnamefont {Matsuyama}}, \bibinfo
  {author} {\bibfnamefont {T.}~\bibnamefont {Koyama}}, \bibinfo {author}
  {\bibfnamefont {K.}~\bibnamefont {Tono}}, \bibinfo {author} {\bibfnamefont
  {Y.}~\bibnamefont {Inubushi}}, \bibinfo {author} {\bibfnamefont
  {T.}~\bibnamefont {Togashi}}, \bibinfo {author} {\bibfnamefont
  {T.}~\bibnamefont {Sato}}, \bibinfo {author} {\bibfnamefont {J.}~\bibnamefont
  {Kim}}, \bibinfo {author} {\bibfnamefont {R.}~\bibnamefont {Fukui}}, \emph
  {et~al.},\ }\bibfield  {title} {\bibinfo {title} {Generation of 10 20 w cm- 2
  hard x-ray laser pulses with two-stage reflective focusing system},\
  }\href@noop {} {\bibfield  {journal} {\bibinfo  {journal} {Nature
  communications}\ }\textbf {\bibinfo {volume} {5}},\ \bibinfo {pages} {1}
  (\bibinfo {year} {2014})}\BibitemShut {NoStop}%
\bibitem [{\citenamefont {Tanaka}\ and\ \citenamefont
  {Mukamel}(2002)}]{tanaka2002coherent}%
  \BibitemOpen
  \bibfield  {author} {\bibinfo {author} {\bibfnamefont {S.}~\bibnamefont
  {Tanaka}}\ and\ \bibinfo {author} {\bibfnamefont {S.}~\bibnamefont
  {Mukamel}},\ }\bibfield  {title} {\bibinfo {title} {Coherent x-ray raman
  spectroscopy: a nonlinear local probe for electronic excitations},\
  }\href@noop {} {\bibfield  {journal} {\bibinfo  {journal} {Physical review
  letters}\ }\textbf {\bibinfo {volume} {89}},\ \bibinfo {pages} {043001}
  (\bibinfo {year} {2002})}\BibitemShut {NoStop}%
\bibitem [{\citenamefont {Schweigert}\ and\ \citenamefont
  {Mukamel}(2008)}]{schweigert2008double}%
  \BibitemOpen
  \bibfield  {author} {\bibinfo {author} {\bibfnamefont {I.~V.}\ \bibnamefont
  {Schweigert}}\ and\ \bibinfo {author} {\bibfnamefont {S.}~\bibnamefont
  {Mukamel}},\ }\bibfield  {title} {\bibinfo {title} {Double-quantum-coherence
  attosecond x-ray spectroscopy of spatially separated, spectrally overlapping
  core-electron transitions},\ }\href@noop {} {\bibfield  {journal} {\bibinfo
  {journal} {Physical Review A}\ }\textbf {\bibinfo {volume} {78}},\ \bibinfo
  {pages} {052509} (\bibinfo {year} {2008})}\BibitemShut {NoStop}%
\bibitem [{\citenamefont {Gr{\"u}bel}\ \emph {et~al.}(2007)\citenamefont
  {Gr{\"u}bel}, \citenamefont {Stephenson}, \citenamefont {Gutt}, \citenamefont
  {Sinn},\ and\ \citenamefont {Tschentscher}}]{Grubel2007}%
  \BibitemOpen
  \bibfield  {author} {\bibinfo {author} {\bibfnamefont {G.}~\bibnamefont
  {Gr{\"u}bel}}, \bibinfo {author} {\bibfnamefont {G.}~\bibnamefont
  {Stephenson}}, \bibinfo {author} {\bibfnamefont {C.}~\bibnamefont {Gutt}},
  \bibinfo {author} {\bibfnamefont {H.}~\bibnamefont {Sinn}},\ and\ \bibinfo
  {author} {\bibfnamefont {T.}~\bibnamefont {Tschentscher}},\ }\bibfield
  {title} {\bibinfo {title} {Xpcs at the european x-ray free electron laser
  facility},\ }\href@noop {} {\bibfield  {journal} {\bibinfo  {journal} {Nucl.
  Instrum. Methods Phys. Res. B}\ }\textbf {\bibinfo {volume} {262}},\ \bibinfo
  {pages} {357} (\bibinfo {year} {2007})}\BibitemShut {NoStop}%
\bibitem [{\citenamefont {Gutt}\ \emph {et~al.}(2009)\citenamefont {Gutt},
  \citenamefont {Stadler}, \citenamefont {Duri}, \citenamefont {Autenrieth},
  \citenamefont {Leupold}, \citenamefont {Chushkin},\ and\ \citenamefont
  {Gr\"{u}bel}}]{Gutt2009}%
  \BibitemOpen
  \bibfield  {author} {\bibinfo {author} {\bibfnamefont {C.}~\bibnamefont
  {Gutt}}, \bibinfo {author} {\bibfnamefont {L.-M.}\ \bibnamefont {Stadler}},
  \bibinfo {author} {\bibfnamefont {A.}~\bibnamefont {Duri}}, \bibinfo {author}
  {\bibfnamefont {T.}~\bibnamefont {Autenrieth}}, \bibinfo {author}
  {\bibfnamefont {O.}~\bibnamefont {Leupold}}, \bibinfo {author} {\bibfnamefont
  {Y.}~\bibnamefont {Chushkin}},\ and\ \bibinfo {author} {\bibfnamefont
  {G.}~\bibnamefont {Gr\"{u}bel}},\ }\bibfield  {title} {\bibinfo {title}
  {Measuring temporal speckle correlations at ultrafast x-ray sources},\ }\href
  {https://doi.org/10.1364/OE.17.000055} {\bibfield  {journal} {\bibinfo
  {journal} {Opt. Express}\ }\textbf {\bibinfo {volume} {17}},\ \bibinfo
  {pages} {55} (\bibinfo {year} {2009})}\BibitemShut {NoStop}%
\bibitem [{\citenamefont {Wang}\ \emph {et~al.}(2019)\citenamefont {Wang},
  \citenamefont {Xu}, \citenamefont {Zhang},\ and\ \citenamefont
  {Douglas}}]{wang2019universal}%
  \BibitemOpen
  \bibfield  {author} {\bibinfo {author} {\bibfnamefont {X.}~\bibnamefont
  {Wang}}, \bibinfo {author} {\bibfnamefont {W.-S.}\ \bibnamefont {Xu}},
  \bibinfo {author} {\bibfnamefont {H.}~\bibnamefont {Zhang}},\ and\ \bibinfo
  {author} {\bibfnamefont {J.~F.}\ \bibnamefont {Douglas}},\ }\bibfield
  {title} {\bibinfo {title} {Universal nature of dynamic heterogeneity in
  glass-forming liquids: A comparative study of metallic and polymeric
  glass-forming liquids},\ }\href@noop {} {\bibfield  {journal} {\bibinfo
  {journal} {The Journal of chemical physics}\ }\textbf {\bibinfo {volume}
  {151}},\ \bibinfo {pages} {184503} (\bibinfo {year} {2019})}\BibitemShut
  {NoStop}%
\bibitem [{\citenamefont {Cicerone}\ \emph {et~al.}(2014)\citenamefont
  {Cicerone}, \citenamefont {Zhong},\ and\ \citenamefont
  {Tyagi}}]{PhysRevLett.113.117801}%
  \BibitemOpen
  \bibfield  {author} {\bibinfo {author} {\bibfnamefont {M.~T.}\ \bibnamefont
  {Cicerone}}, \bibinfo {author} {\bibfnamefont {Q.}~\bibnamefont {Zhong}},\
  and\ \bibinfo {author} {\bibfnamefont {M.}~\bibnamefont {Tyagi}},\ }\bibfield
   {title} {\bibinfo {title} {Picosecond dynamic heterogeneity, hopping, and
  johari-goldstein relaxation in glass-forming liquids},\ }\href
  {https://doi.org/10.1103/PhysRevLett.113.117801} {\bibfield  {journal}
  {\bibinfo  {journal} {Phys. Rev. Lett.}\ }\textbf {\bibinfo {volume} {113}},\
  \bibinfo {pages} {117801} (\bibinfo {year} {2014})}\BibitemShut {NoStop}%
\bibitem [{\citenamefont {Roseker}\ \emph {et~al.}(2009)\citenamefont
  {Roseker}, \citenamefont {Franz}, \citenamefont {Schulte-Schrepping},
  \citenamefont {Ehnes}, \citenamefont {Leupold}, \citenamefont {Zontone},
  \citenamefont {Robert},\ and\ \citenamefont
  {Gr{\"u}bel}}]{roseker2009performance}%
  \BibitemOpen
  \bibfield  {author} {\bibinfo {author} {\bibfnamefont {W.}~\bibnamefont
  {Roseker}}, \bibinfo {author} {\bibfnamefont {H.}~\bibnamefont {Franz}},
  \bibinfo {author} {\bibfnamefont {H.}~\bibnamefont {Schulte-Schrepping}},
  \bibinfo {author} {\bibfnamefont {A.}~\bibnamefont {Ehnes}}, \bibinfo
  {author} {\bibfnamefont {O.}~\bibnamefont {Leupold}}, \bibinfo {author}
  {\bibfnamefont {F.}~\bibnamefont {Zontone}}, \bibinfo {author} {\bibfnamefont
  {A.}~\bibnamefont {Robert}},\ and\ \bibinfo {author} {\bibfnamefont
  {G.}~\bibnamefont {Gr{\"u}bel}},\ }\bibfield  {title} {\bibinfo {title}
  {Performance of a picosecond x-ray delay line unit at 8.39 kev},\ }\href@noop
  {} {\bibfield  {journal} {\bibinfo  {journal} {Optics letters}\ }\textbf
  {\bibinfo {volume} {34}},\ \bibinfo {pages} {1768} (\bibinfo {year}
  {2009})}\BibitemShut {NoStop}%
\bibitem [{\citenamefont {Osaka}\ \emph {et~al.}(2013)\citenamefont {Osaka},
  \citenamefont {Yabashi}, \citenamefont {Sano}, \citenamefont {Tono},
  \citenamefont {Inubushi}, \citenamefont {Sato}, \citenamefont {Matsuyama},
  \citenamefont {Ishikawa},\ and\ \citenamefont {Yamauchi}}]{osaka2013bragg}%
  \BibitemOpen
  \bibfield  {author} {\bibinfo {author} {\bibfnamefont {T.}~\bibnamefont
  {Osaka}}, \bibinfo {author} {\bibfnamefont {M.}~\bibnamefont {Yabashi}},
  \bibinfo {author} {\bibfnamefont {Y.}~\bibnamefont {Sano}}, \bibinfo {author}
  {\bibfnamefont {K.}~\bibnamefont {Tono}}, \bibinfo {author} {\bibfnamefont
  {Y.}~\bibnamefont {Inubushi}}, \bibinfo {author} {\bibfnamefont
  {T.}~\bibnamefont {Sato}}, \bibinfo {author} {\bibfnamefont {S.}~\bibnamefont
  {Matsuyama}}, \bibinfo {author} {\bibfnamefont {T.}~\bibnamefont
  {Ishikawa}},\ and\ \bibinfo {author} {\bibfnamefont {K.}~\bibnamefont
  {Yamauchi}},\ }\bibfield  {title} {\bibinfo {title} {A bragg beam splitter
  for hard x-ray free-electron lasers},\ }\href@noop {} {\bibfield  {journal}
  {\bibinfo  {journal} {Optics express}\ }\textbf {\bibinfo {volume} {21}},\
  \bibinfo {pages} {2823} (\bibinfo {year} {2013})}\BibitemShut {NoStop}%
\bibitem [{\citenamefont {Lu}\ \emph {et~al.}(2018)\citenamefont {Lu},
  \citenamefont {Friedrich}, \citenamefont {Noll}, \citenamefont {Zhou},
  \citenamefont {Hallmann}, \citenamefont {Ansaldi}, \citenamefont {Roth},
  \citenamefont {Serkez}, \citenamefont {Geloni}, \citenamefont {Madsen},\ and\
  \citenamefont {Eisebitt}}]{lu2018development}%
  \BibitemOpen
  \bibfield  {author} {\bibinfo {author} {\bibfnamefont {W.}~\bibnamefont
  {Lu}}, \bibinfo {author} {\bibfnamefont {B.}~\bibnamefont {Friedrich}},
  \bibinfo {author} {\bibfnamefont {T.}~\bibnamefont {Noll}}, \bibinfo {author}
  {\bibfnamefont {K.}~\bibnamefont {Zhou}}, \bibinfo {author} {\bibfnamefont
  {J.}~\bibnamefont {Hallmann}}, \bibinfo {author} {\bibfnamefont
  {G.}~\bibnamefont {Ansaldi}}, \bibinfo {author} {\bibfnamefont
  {T.}~\bibnamefont {Roth}}, \bibinfo {author} {\bibfnamefont {S.}~\bibnamefont
  {Serkez}}, \bibinfo {author} {\bibfnamefont {G.}~\bibnamefont {Geloni}},
  \bibinfo {author} {\bibfnamefont {A.}~\bibnamefont {Madsen}},\ and\ \bibinfo
  {author} {\bibfnamefont {S.}~\bibnamefont {Eisebitt}},\ }\bibfield  {title}
  {\bibinfo {title} {Development of a hard x-ray split-and-delay line and
  performance simulations for two-color pump-probe experiments at the european
  xfel},\ }\href@noop {} {\bibfield  {journal} {\bibinfo  {journal} {Review of
  Scientific Instruments}\ }\textbf {\bibinfo {volume} {89}},\ \bibinfo {pages}
  {063121} (\bibinfo {year} {2018})}\BibitemShut {NoStop}%
\bibitem [{\citenamefont {Rysov}\ \emph {et~al.}(2019)\citenamefont {Rysov},
  \citenamefont {Roseker}, \citenamefont {Walther},\ and\ \citenamefont
  {Gr{\"u}bel}}]{rysov2019compact}%
  \BibitemOpen
  \bibfield  {author} {\bibinfo {author} {\bibfnamefont {R.}~\bibnamefont
  {Rysov}}, \bibinfo {author} {\bibfnamefont {W.}~\bibnamefont {Roseker}},
  \bibinfo {author} {\bibfnamefont {M.}~\bibnamefont {Walther}},\ and\ \bibinfo
  {author} {\bibfnamefont {G.}~\bibnamefont {Gr{\"u}bel}},\ }\bibfield  {title}
  {\bibinfo {title} {Compact hard x-ray split-and-delay line for studying
  ultrafast dynamics at free-electron laser sources},\ }\href@noop {}
  {\bibfield  {journal} {\bibinfo  {journal} {Journal of Synchrotron
  Radiation}\ }\textbf {\bibinfo {volume} {26}} (\bibinfo {year}
  {2019})}\BibitemShut {NoStop}%
\bibitem [{\citenamefont {Osaka}\ \emph {et~al.}(2016)\citenamefont {Osaka},
  \citenamefont {Hirano}, \citenamefont {Sano}, \citenamefont {Inubushi},
  \citenamefont {Matsuyama}, \citenamefont {Tono}, \citenamefont {Ishikawa},
  \citenamefont {Yamauchi},\ and\ \citenamefont
  {Yabashi}}]{osaka2016wavelength}%
  \BibitemOpen
  \bibfield  {author} {\bibinfo {author} {\bibfnamefont {T.}~\bibnamefont
  {Osaka}}, \bibinfo {author} {\bibfnamefont {T.}~\bibnamefont {Hirano}},
  \bibinfo {author} {\bibfnamefont {Y.}~\bibnamefont {Sano}}, \bibinfo {author}
  {\bibfnamefont {Y.}~\bibnamefont {Inubushi}}, \bibinfo {author}
  {\bibfnamefont {S.}~\bibnamefont {Matsuyama}}, \bibinfo {author}
  {\bibfnamefont {K.}~\bibnamefont {Tono}}, \bibinfo {author} {\bibfnamefont
  {T.}~\bibnamefont {Ishikawa}}, \bibinfo {author} {\bibfnamefont
  {K.}~\bibnamefont {Yamauchi}},\ and\ \bibinfo {author} {\bibfnamefont
  {M.}~\bibnamefont {Yabashi}},\ }\bibfield  {title} {\bibinfo {title}
  {Wavelength-tunable split-and-delay optical system for hard x-ray
  free-electron lasers},\ }\href@noop {} {\bibfield  {journal} {\bibinfo
  {journal} {Optics express}\ }\textbf {\bibinfo {volume} {24}},\ \bibinfo
  {pages} {9187} (\bibinfo {year} {2016})}\BibitemShut {NoStop}%
\bibitem [{\citenamefont {Zhu}\ \emph {et~al.}(2017)\citenamefont {Zhu},
  \citenamefont {Sun}, \citenamefont {Schafer}, \citenamefont {Shi},
  \citenamefont {James}, \citenamefont {Gumerlock}, \citenamefont {Osier},
  \citenamefont {Whitney}, \citenamefont {Zhang}, \citenamefont {Nicolas},
  \citenamefont {Smith}, \citenamefont {Barada},\ and\ \citenamefont
  {Robert}}]{zhu2017development}%
  \BibitemOpen
  \bibfield  {author} {\bibinfo {author} {\bibfnamefont {D.}~\bibnamefont
  {Zhu}}, \bibinfo {author} {\bibfnamefont {Y.}~\bibnamefont {Sun}}, \bibinfo
  {author} {\bibfnamefont {D.~W.}\ \bibnamefont {Schafer}}, \bibinfo {author}
  {\bibfnamefont {H.}~\bibnamefont {Shi}}, \bibinfo {author} {\bibfnamefont
  {J.~H.}\ \bibnamefont {James}}, \bibinfo {author} {\bibfnamefont {K.~L.}\
  \bibnamefont {Gumerlock}}, \bibinfo {author} {\bibfnamefont {T.~O.}\
  \bibnamefont {Osier}}, \bibinfo {author} {\bibfnamefont {R.}~\bibnamefont
  {Whitney}}, \bibinfo {author} {\bibfnamefont {L.}~\bibnamefont {Zhang}},
  \bibinfo {author} {\bibfnamefont {J.}~\bibnamefont {Nicolas}}, \bibinfo
  {author} {\bibfnamefont {B.}~\bibnamefont {Smith}}, \bibinfo {author}
  {\bibfnamefont {A.~H.}\ \bibnamefont {Barada}},\ and\ \bibinfo {author}
  {\bibfnamefont {A.}~\bibnamefont {Robert}},\ }\bibfield  {title} {\bibinfo
  {title} {Development of a hard x-ray split-delay system at the linac coherent
  light source},\ }\href@noop {} {\bibfield  {journal} {\bibinfo  {journal}
  {Advances in X-ray Free-Electron Lasers Instrumentation IV}\ }\textbf
  {\bibinfo {volume} {10237}},\ \bibinfo {pages} {102370R} (\bibinfo {year}
  {2017})}\BibitemShut {NoStop}%
\bibitem [{\citenamefont {Shi}\ and\ \citenamefont {Zhu}(2018)}]{shi2018multi}%
  \BibitemOpen
  \bibfield  {author} {\bibinfo {author} {\bibfnamefont {H.}~\bibnamefont
  {Shi}}\ and\ \bibinfo {author} {\bibfnamefont {D.}~\bibnamefont {Zhu}},\
  }\bibfield  {title} {\bibinfo {title} {Multi-axis nanopositioning system for
  the hard x-ray split-delay system at the lcls},\ }\href@noop {} {\bibfield
  {journal} {\bibinfo  {journal} {Synchrotron Radiation News}\ }\textbf
  {\bibinfo {volume} {31}},\ \bibinfo {pages} {15} (\bibinfo {year}
  {2018})}\BibitemShut {NoStop}%
\bibitem [{\citenamefont {Sun}\ \emph {et~al.}(2019{\natexlab{a}})\citenamefont
  {Sun}, \citenamefont {Wang}, \citenamefont {Song}, \citenamefont {Sun},
  \citenamefont {Chollet}, \citenamefont {Sato}, \citenamefont {van Driel},
  \citenamefont {Nelson}, \citenamefont {Plumley}, \citenamefont
  {Montana-Lopez}, \citenamefont {Teitelbaum}, \citenamefont {Haber},
  \citenamefont {Hastings}, \citenamefont {Baron}, \citenamefont {Sutton},
  \citenamefont {Fuoss}, \citenamefont {Robert},\ and\ \citenamefont
  {Zhu}}]{sun2019compact}%
  \BibitemOpen
  \bibfield  {author} {\bibinfo {author} {\bibfnamefont {Y.}~\bibnamefont
  {Sun}}, \bibinfo {author} {\bibfnamefont {N.}~\bibnamefont {Wang}}, \bibinfo
  {author} {\bibfnamefont {S.}~\bibnamefont {Song}}, \bibinfo {author}
  {\bibfnamefont {P.}~\bibnamefont {Sun}}, \bibinfo {author} {\bibfnamefont
  {M.}~\bibnamefont {Chollet}}, \bibinfo {author} {\bibfnamefont
  {T.}~\bibnamefont {Sato}}, \bibinfo {author} {\bibfnamefont {T.~B.}\
  \bibnamefont {van Driel}}, \bibinfo {author} {\bibfnamefont {S.}~\bibnamefont
  {Nelson}}, \bibinfo {author} {\bibfnamefont {R.}~\bibnamefont {Plumley}},
  \bibinfo {author} {\bibfnamefont {J.}~\bibnamefont {Montana-Lopez}}, \bibinfo
  {author} {\bibfnamefont {S.~W.}\ \bibnamefont {Teitelbaum}}, \bibinfo
  {author} {\bibfnamefont {J.}~\bibnamefont {Haber}}, \bibinfo {author}
  {\bibfnamefont {J.~B.}\ \bibnamefont {Hastings}}, \bibinfo {author}
  {\bibfnamefont {A.~Q.~R.}\ \bibnamefont {Baron}}, \bibinfo {author}
  {\bibfnamefont {M.}~\bibnamefont {Sutton}}, \bibinfo {author} {\bibfnamefont
  {P.~H.}\ \bibnamefont {Fuoss}}, \bibinfo {author} {\bibfnamefont
  {A.}~\bibnamefont {Robert}},\ and\ \bibinfo {author} {\bibfnamefont
  {D.}~\bibnamefont {Zhu}},\ }\bibfield  {title} {\bibinfo {title} {Compact
  hard x-ray split-delay system based on variable-gap channel-cut crystals},\
  }\href@noop {} {\bibfield  {journal} {\bibinfo  {journal} {Opt. Lett.}\
  }\textbf {\bibinfo {volume} {44}},\ \bibinfo {pages} {2582} (\bibinfo {year}
  {2019}{\natexlab{a}})}\BibitemShut {NoStop}%
\bibitem [{\citenamefont {Sun}\ \emph {et~al.}(2019{\natexlab{b}})\citenamefont
  {Sun}, \citenamefont {Robert},\ and\ \citenamefont {Zhu}}]{sun2019design}%
  \BibitemOpen
  \bibfield  {author} {\bibinfo {author} {\bibfnamefont {Y.}~\bibnamefont
  {Sun}}, \bibinfo {author} {\bibfnamefont {A.}~\bibnamefont {Robert}},\ and\
  \bibinfo {author} {\bibfnamefont {D.}~\bibnamefont {Zhu}},\ }\bibfield
  {title} {\bibinfo {title} {Design of a compact hard x-ray split-delay system
  based on variable-gap channelcut crystals},\ }\href@noop {} {\bibfield
  {journal} {\bibinfo  {journal} {AIP Conference Proceedings}\ }\textbf
  {\bibinfo {volume} {2054}},\ \bibinfo {pages} {060004} (\bibinfo {year}
  {2019}{\natexlab{b}})}\BibitemShut {NoStop}%
\bibitem [{\citenamefont {Sun}\ \emph {et~al.}(2020{\natexlab{a}})\citenamefont
  {Sun}, \citenamefont {Dunne}, \citenamefont {Fuoss}, \citenamefont {Robert},
  \citenamefont {Zhu}, \citenamefont {Osaka}, \citenamefont {Yabashi},\ and\
  \citenamefont {Sutton}}]{sun2020realizing}%
  \BibitemOpen
  \bibfield  {author} {\bibinfo {author} {\bibfnamefont {Y.}~\bibnamefont
  {Sun}}, \bibinfo {author} {\bibfnamefont {M.}~\bibnamefont {Dunne}}, \bibinfo
  {author} {\bibfnamefont {P.}~\bibnamefont {Fuoss}}, \bibinfo {author}
  {\bibfnamefont {A.}~\bibnamefont {Robert}}, \bibinfo {author} {\bibfnamefont
  {D.}~\bibnamefont {Zhu}}, \bibinfo {author} {\bibfnamefont {T.}~\bibnamefont
  {Osaka}}, \bibinfo {author} {\bibfnamefont {M.}~\bibnamefont {Yabashi}},\
  and\ \bibinfo {author} {\bibfnamefont {M.}~\bibnamefont {Sutton}},\
  }\bibfield  {title} {\bibinfo {title} {Realizing split-pulse x-ray photon
  correlation spectroscopy to measure ultrafast dynamics in complex matter},\
  }\href@noop {} {\bibfield  {journal} {\bibinfo  {journal} {Physical Review
  Research}\ }\textbf {\bibinfo {volume} {2}},\ \bibinfo {pages} {023099}
  (\bibinfo {year} {2020}{\natexlab{a}})}\BibitemShut {NoStop}%
\bibitem [{\citenamefont {Li}\ \emph {et~al.}(2020{\natexlab{a}})\citenamefont
  {Li}, \citenamefont {Sun}, \citenamefont {Sutton}, \citenamefont {Fuoss},\
  and\ \citenamefont {Zhu}}]{li2020design}%
  \BibitemOpen
  \bibfield  {author} {\bibinfo {author} {\bibfnamefont {H.}~\bibnamefont
  {Li}}, \bibinfo {author} {\bibfnamefont {Y.}~\bibnamefont {Sun}}, \bibinfo
  {author} {\bibfnamefont {M.}~\bibnamefont {Sutton}}, \bibinfo {author}
  {\bibfnamefont {P.}~\bibnamefont {Fuoss}},\ and\ \bibinfo {author}
  {\bibfnamefont {D.}~\bibnamefont {Zhu}},\ }\bibfield  {title} {\bibinfo
  {title} {Design of an amplitude-splitting hard x-ray delay line with
  subnanoradian stability},\ }\href@noop {} {\bibfield  {journal} {\bibinfo
  {journal} {Optics Letters}\ }\textbf {\bibinfo {volume} {45}},\ \bibinfo
  {pages} {2086} (\bibinfo {year} {2020}{\natexlab{a}})}\BibitemShut {NoStop}%
\bibitem [{\citenamefont {Chollet}\ \emph {et~al.}(2015)\citenamefont
  {Chollet}, \citenamefont {Alonso-Mori}, \citenamefont {Cammarata},
  \citenamefont {Damiani}, \citenamefont {Defever}, \citenamefont {Delor},
  \citenamefont {Feng}, \citenamefont {Glownia}, \citenamefont {Langton},
  \citenamefont {Nelson}, \citenamefont {Ramsey}, \citenamefont {Robert},
  \citenamefont {Sikorski}, \citenamefont {Song}, \citenamefont {Stefanescu},
  \citenamefont {Srinivasan}, \citenamefont {Zhu}, \citenamefont {Lemke},\ and\
  \citenamefont {Fritz}}]{Chollet2015}%
  \BibitemOpen
  \bibfield  {author} {\bibinfo {author} {\bibfnamefont {M.}~\bibnamefont
  {Chollet}}, \bibinfo {author} {\bibfnamefont {R.}~\bibnamefont
  {Alonso-Mori}}, \bibinfo {author} {\bibfnamefont {M.}~\bibnamefont
  {Cammarata}}, \bibinfo {author} {\bibfnamefont {D.}~\bibnamefont {Damiani}},
  \bibinfo {author} {\bibfnamefont {J.}~\bibnamefont {Defever}}, \bibinfo
  {author} {\bibfnamefont {J.~T.}\ \bibnamefont {Delor}}, \bibinfo {author}
  {\bibfnamefont {Y.}~\bibnamefont {Feng}}, \bibinfo {author} {\bibfnamefont
  {J.~M.}\ \bibnamefont {Glownia}}, \bibinfo {author} {\bibfnamefont {J.~B.}\
  \bibnamefont {Langton}}, \bibinfo {author} {\bibfnamefont {S.}~\bibnamefont
  {Nelson}}, \bibinfo {author} {\bibfnamefont {K.}~\bibnamefont {Ramsey}},
  \bibinfo {author} {\bibfnamefont {A.}~\bibnamefont {Robert}}, \bibinfo
  {author} {\bibfnamefont {M.}~\bibnamefont {Sikorski}}, \bibinfo {author}
  {\bibfnamefont {S.}~\bibnamefont {Song}}, \bibinfo {author} {\bibfnamefont
  {D.}~\bibnamefont {Stefanescu}}, \bibinfo {author} {\bibfnamefont
  {V.}~\bibnamefont {Srinivasan}}, \bibinfo {author} {\bibfnamefont
  {D.}~\bibnamefont {Zhu}}, \bibinfo {author} {\bibfnamefont {H.~T.}\
  \bibnamefont {Lemke}},\ and\ \bibinfo {author} {\bibfnamefont {D.~M.}\
  \bibnamefont {Fritz}},\ }\bibfield  {title} {\bibinfo {title} {{The {X}-ray
  {P}ump{--}{P}robe instrument at the {Linac~Coherent Light Source}}},\
  }\href@noop {} {\bibfield  {journal} {\bibinfo  {journal} {Journal of
  Synchrotron Radiation}\ }\textbf {\bibinfo {volume} {22}},\ \bibinfo {pages}
  {503} (\bibinfo {year} {2015})}\BibitemShut {NoStop}%
\bibitem [{\citenamefont {Zhu}\ \emph {et~al.}(2014)\citenamefont {Zhu},
  \citenamefont {Feng}, \citenamefont {Stoupin}, \citenamefont {Terentyev},
  \citenamefont {Lemke}, \citenamefont {Fritz}, \citenamefont {Chollet},
  \citenamefont {Glownia}, \citenamefont {Alonso-Mori}, \citenamefont
  {Sikorski} \emph {et~al.}}]{zhu2014performance}%
  \BibitemOpen
  \bibfield  {author} {\bibinfo {author} {\bibfnamefont {D.}~\bibnamefont
  {Zhu}}, \bibinfo {author} {\bibfnamefont {Y.}~\bibnamefont {Feng}}, \bibinfo
  {author} {\bibfnamefont {S.}~\bibnamefont {Stoupin}}, \bibinfo {author}
  {\bibfnamefont {S.~A.}\ \bibnamefont {Terentyev}}, \bibinfo {author}
  {\bibfnamefont {H.~T.}\ \bibnamefont {Lemke}}, \bibinfo {author}
  {\bibfnamefont {D.~M.}\ \bibnamefont {Fritz}}, \bibinfo {author}
  {\bibfnamefont {M.}~\bibnamefont {Chollet}}, \bibinfo {author} {\bibfnamefont
  {J.}~\bibnamefont {Glownia}}, \bibinfo {author} {\bibfnamefont
  {R.}~\bibnamefont {Alonso-Mori}}, \bibinfo {author} {\bibfnamefont
  {M.}~\bibnamefont {Sikorski}}, \emph {et~al.},\ }\bibfield  {title} {\bibinfo
  {title} {Performance of a beam-multiplexing diamond crystal monochromator at
  the linac coherent light source},\ }\href@noop {} {\bibfield  {journal}
  {\bibinfo  {journal} {Review of Scientific Instruments}\ }\textbf {\bibinfo
  {volume} {85}},\ \bibinfo {pages} {063106} (\bibinfo {year}
  {2014})}\BibitemShut {NoStop}%
\bibitem [{\citenamefont {Makita}\ \emph {et~al.}(2017)\citenamefont {Makita},
  \citenamefont {Karvinen}, \citenamefont {Guzenko}, \citenamefont {Kujala},
  \citenamefont {Vagovic},\ and\ \citenamefont {David}}]{Mikata2017}%
  \BibitemOpen
  \bibfield  {author} {\bibinfo {author} {\bibfnamefont {M.}~\bibnamefont
  {Makita}}, \bibinfo {author} {\bibfnamefont {P.}~\bibnamefont {Karvinen}},
  \bibinfo {author} {\bibfnamefont {V.}~\bibnamefont {Guzenko}}, \bibinfo
  {author} {\bibfnamefont {N.}~\bibnamefont {Kujala}}, \bibinfo {author}
  {\bibfnamefont {P.}~\bibnamefont {Vagovic}},\ and\ \bibinfo {author}
  {\bibfnamefont {C.}~\bibnamefont {David}},\ }\bibfield  {title} {\bibinfo
  {title} {Fabrication of diamond diffraction gratings for experiments with
  intense hard x-rays},\ }\href
  {https://doi.org/https://doi.org/10.1016/j.mee.2017.02.002} {\bibfield
  {journal} {\bibinfo  {journal} {Microelectronic Engineering}\ }\textbf
  {\bibinfo {volume} {176}},\ \bibinfo {pages} {75} (\bibinfo {year} {2017})},\
  \bibinfo {note} {micro- and Nano-Fabrication}\BibitemShut {NoStop}%
\bibitem [{\citenamefont {Li}\ \emph {et~al.}(2020{\natexlab{b}})\citenamefont
  {Li}, \citenamefont {Liu}, \citenamefont {Seaberg}, \citenamefont {Chollet},
  \citenamefont {Weiss},\ and\ \citenamefont {Sakdinawat}}]{Li2020grating}%
  \BibitemOpen
  \bibfield  {author} {\bibinfo {author} {\bibfnamefont {K.}~\bibnamefont
  {Li}}, \bibinfo {author} {\bibfnamefont {Y.}~\bibnamefont {Liu}}, \bibinfo
  {author} {\bibfnamefont {M.}~\bibnamefont {Seaberg}}, \bibinfo {author}
  {\bibfnamefont {M.}~\bibnamefont {Chollet}}, \bibinfo {author} {\bibfnamefont
  {T.~M.}\ \bibnamefont {Weiss}},\ and\ \bibinfo {author} {\bibfnamefont
  {A.}~\bibnamefont {Sakdinawat}},\ }\bibfield  {title} {\bibinfo {title}
  {Wavefront preserving and high efficiency diamond grating beam splitter for
  x-ray free electron laser},\ }\href {https://doi.org/10.1364/OE.380534}
  {\bibfield  {journal} {\bibinfo  {journal} {Opt. Express}\ }\textbf {\bibinfo
  {volume} {28}},\ \bibinfo {pages} {10939} (\bibinfo {year}
  {2020}{\natexlab{b}})}\BibitemShut {NoStop}%
\bibitem [{\citenamefont {{Carini}}\ \emph {et~al.}(2014)\citenamefont
  {{Carini}}, \citenamefont {{Alonso-Mori}}, \citenamefont {{Blaj}},
  \citenamefont {{Caragiulo}}, \citenamefont {{Chollet}}, \citenamefont
  {{Damiani}}, \citenamefont {{Dragone}}, \citenamefont {{Feng}}, \citenamefont
  {{Haller}}, \citenamefont {{Hart}}, \citenamefont {{Hasi}}, \citenamefont
  {{Herbst}}, \citenamefont {{Herrmann}}, \citenamefont {{Kenney}},
  \citenamefont {{Lemke}}, \citenamefont {{Markovic}}, \citenamefont
  {{Nelson}}, \citenamefont {{Nishimura}}, \citenamefont {{Osier}},
  \citenamefont {{Pines}}, \citenamefont {{Robert}}, \citenamefont {{Segal}},
  \citenamefont {{Sikorski}}, \citenamefont {{Song}}, \citenamefont {{Tomada}},
  \citenamefont {{Weaver}},\ and\ \citenamefont {Zhu}}]{Carini2014}%
  \BibitemOpen
  \bibfield  {author} {\bibinfo {author} {\bibfnamefont {G.~A.}\ \bibnamefont
  {{Carini}}}, \bibinfo {author} {\bibfnamefont {R.}~\bibnamefont
  {{Alonso-Mori}}}, \bibinfo {author} {\bibfnamefont {G.}~\bibnamefont
  {{Blaj}}}, \bibinfo {author} {\bibfnamefont {P.}~\bibnamefont {{Caragiulo}}},
  \bibinfo {author} {\bibfnamefont {M.}~\bibnamefont {{Chollet}}}, \bibinfo
  {author} {\bibfnamefont {D.}~\bibnamefont {{Damiani}}}, \bibinfo {author}
  {\bibfnamefont {A.}~\bibnamefont {{Dragone}}}, \bibinfo {author}
  {\bibfnamefont {Y.}~\bibnamefont {{Feng}}}, \bibinfo {author} {\bibfnamefont
  {G.}~\bibnamefont {{Haller}}}, \bibinfo {author} {\bibfnamefont
  {P.}~\bibnamefont {{Hart}}}, \bibinfo {author} {\bibfnamefont
  {J.}~\bibnamefont {{Hasi}}}, \bibinfo {author} {\bibfnamefont
  {R.}~\bibnamefont {{Herbst}}}, \bibinfo {author} {\bibfnamefont
  {S.}~\bibnamefont {{Herrmann}}}, \bibinfo {author} {\bibfnamefont
  {C.}~\bibnamefont {{Kenney}}}, \bibinfo {author} {\bibfnamefont
  {H.}~\bibnamefont {{Lemke}}}, \bibinfo {author} {\bibfnamefont
  {B.}~\bibnamefont {{Markovic}}}, \bibinfo {author} {\bibfnamefont
  {S.}~\bibnamefont {{Nelson}}}, \bibinfo {author} {\bibfnamefont
  {K.}~\bibnamefont {{Nishimura}}}, \bibinfo {author} {\bibfnamefont
  {S.}~\bibnamefont {{Osier}}}, \bibinfo {author} {\bibfnamefont
  {J.}~\bibnamefont {{Pines}}}, \bibinfo {author} {\bibfnamefont
  {A.}~\bibnamefont {{Robert}}}, \bibinfo {author} {\bibfnamefont
  {J.}~\bibnamefont {{Segal}}}, \bibinfo {author} {\bibfnamefont
  {M.}~\bibnamefont {{Sikorski}}}, \bibinfo {author} {\bibfnamefont
  {S.}~\bibnamefont {{Song}}}, \bibinfo {author} {\bibfnamefont
  {A.}~\bibnamefont {{Tomada}}}, \bibinfo {author} {\bibfnamefont
  {M.}~\bibnamefont {{Weaver}}},\ and\ \bibinfo {author} {\bibfnamefont
  {D.}~\bibnamefont {Zhu}},\ }\bibfield  {title} {\bibinfo {title} {Studies of
  the {ePix100} low-noise x-ray camera at {SLAC}},\ }in\ \href
  {https://doi.org/10.1109/NSSMIC.2014.7431079} {\emph {\bibinfo {booktitle}
  {2014 IEEE Nuclear Science Symposium and Medical Imaging Conference
  (NSS/MIC)}}}\ (\bibinfo {year} {2014})\ pp.\ \bibinfo {pages}
  {1--3}\BibitemShut {NoStop}%
\bibitem [{\citenamefont {Osaka}\ \emph {et~al.}(2017)\citenamefont {Osaka},
  \citenamefont {Hirano}, \citenamefont {Morioka}, \citenamefont {Sano},
  \citenamefont {Inubushi}, \citenamefont {Togashi}, \citenamefont {Inoue},
  \citenamefont {Tono}, \citenamefont {Robert}, \citenamefont {Yamauchi},
  \citenamefont {Hastings},\ and\ \citenamefont {Yabashi}}]{Osaka2017}%
  \BibitemOpen
  \bibfield  {author} {\bibinfo {author} {\bibfnamefont {T.}~\bibnamefont
  {Osaka}}, \bibinfo {author} {\bibfnamefont {T.}~\bibnamefont {Hirano}},
  \bibinfo {author} {\bibfnamefont {Y.}~\bibnamefont {Morioka}}, \bibinfo
  {author} {\bibfnamefont {Y.}~\bibnamefont {Sano}}, \bibinfo {author}
  {\bibfnamefont {Y.}~\bibnamefont {Inubushi}}, \bibinfo {author}
  {\bibfnamefont {T.}~\bibnamefont {Togashi}}, \bibinfo {author} {\bibfnamefont
  {I.}~\bibnamefont {Inoue}}, \bibinfo {author} {\bibfnamefont
  {K.}~\bibnamefont {Tono}}, \bibinfo {author} {\bibfnamefont {A.}~\bibnamefont
  {Robert}}, \bibinfo {author} {\bibfnamefont {K.}~\bibnamefont {Yamauchi}},
  \bibinfo {author} {\bibfnamefont {J.~B.}\ \bibnamefont {Hastings}},\ and\
  \bibinfo {author} {\bibfnamefont {M.}~\bibnamefont {Yabashi}},\ }\bibfield
  {title} {\bibinfo {title} {{Characterization of temporal coherence of hard
  X-ray free-electron laser pulses with single-shot interferograms}},\
  }\href@noop {} {\bibfield  {journal} {\bibinfo  {journal} {IUCrJ}\ }\textbf
  {\bibinfo {volume} {4}},\ \bibinfo {pages} {728} (\bibinfo {year}
  {2017})}\BibitemShut {NoStop}%
\bibitem [{\citenamefont {Li}(2021)}]{code}%
  \BibitemOpen
  \bibfield  {author} {\bibinfo {author} {\bibfnamefont {H.}~\bibnamefont
  {Li}},\ }\href {https://doi.org/10.6084/m9.figshare.14388437.v2} {\bibinfo
  {title} {Splitdelaysimulation}} (\bibinfo {year} {2021})\BibitemShut
  {NoStop}%
\bibitem [{\citenamefont {Feng}\ \emph {et~al.}(2011)\citenamefont {Feng},
  \citenamefont {Feldkamp}, \citenamefont {Fritz}, \citenamefont {Cammarata},
  \citenamefont {Aymeric}, \citenamefont {Caronna}, \citenamefont {Lemke},
  \citenamefont {\textbf{D. Zhu}}, \citenamefont {Lee}, \citenamefont {Boutet}
  \emph {et~al.}}]{Feng2011}%
  \BibitemOpen
  \bibfield  {author} {\bibinfo {author} {\bibfnamefont {Y.}~\bibnamefont
  {Feng}}, \bibinfo {author} {\bibfnamefont {J.~M.}\ \bibnamefont {Feldkamp}},
  \bibinfo {author} {\bibfnamefont {D.~M.}\ \bibnamefont {Fritz}}, \bibinfo
  {author} {\bibfnamefont {M.}~\bibnamefont {Cammarata}}, \bibinfo {author}
  {\bibfnamefont {R.}~\bibnamefont {Aymeric}}, \bibinfo {author} {\bibfnamefont
  {C.}~\bibnamefont {Caronna}}, \bibinfo {author} {\bibfnamefont {H.~T.}\
  \bibnamefont {Lemke}}, \bibinfo {author} {\bibnamefont {\textbf{D. Zhu}}},
  \bibinfo {author} {\bibfnamefont {S.}~\bibnamefont {Lee}}, \bibinfo {author}
  {\bibfnamefont {S.}~\bibnamefont {Boutet}}, \emph {et~al.},\ }\bibfield
  {title} {\bibinfo {title} {A single-shot intensity-position monitor for hard
  x-ray fel sources},\ }in\ \href@noop {} {\emph {\bibinfo {booktitle} {X-ray
  Lasers and Coherent X-ray Sources: Development and Applications IX}}},\ Vol.\
  \bibinfo {volume} {8140}\ (\bibinfo {organization} {International Society for
  Optics and Photonics},\ \bibinfo {year} {2011})\ p.\ \bibinfo {pages}
  {81400Q}\BibitemShut {NoStop}%
\bibitem [{\citenamefont {Gutt}\ \emph {et~al.}(2012)\citenamefont {Gutt},
  \citenamefont {Wochner}, \citenamefont {Fischer}, \citenamefont {Conrad},
  \citenamefont {Castro-Colin}, \citenamefont {Lee}, \citenamefont
  {Lehmk\"uhler}, \citenamefont {Steinke}, \citenamefont {Sprung},
  \citenamefont {Roseker}, \citenamefont {Zhu}, \citenamefont {Lemke},
  \citenamefont {Bogle}, \citenamefont {Fuoss}, \citenamefont {Stephenson},
  \citenamefont {Cammarata}, \citenamefont {Fritz}, \citenamefont {Robert},\
  and\ \citenamefont {Gr\"ubel}}]{Gutt2012}%
  \BibitemOpen
  \bibfield  {author} {\bibinfo {author} {\bibfnamefont {C.}~\bibnamefont
  {Gutt}}, \bibinfo {author} {\bibfnamefont {P.}~\bibnamefont {Wochner}},
  \bibinfo {author} {\bibfnamefont {B.}~\bibnamefont {Fischer}}, \bibinfo
  {author} {\bibfnamefont {H.}~\bibnamefont {Conrad}}, \bibinfo {author}
  {\bibfnamefont {M.}~\bibnamefont {Castro-Colin}}, \bibinfo {author}
  {\bibfnamefont {S.}~\bibnamefont {Lee}}, \bibinfo {author} {\bibfnamefont
  {F.}~\bibnamefont {Lehmk\"uhler}}, \bibinfo {author} {\bibfnamefont
  {I.}~\bibnamefont {Steinke}}, \bibinfo {author} {\bibfnamefont
  {M.}~\bibnamefont {Sprung}}, \bibinfo {author} {\bibfnamefont
  {W.}~\bibnamefont {Roseker}}, \bibinfo {author} {\bibfnamefont
  {D.}~\bibnamefont {Zhu}}, \bibinfo {author} {\bibfnamefont {H.}~\bibnamefont
  {Lemke}}, \bibinfo {author} {\bibfnamefont {S.}~\bibnamefont {Bogle}},
  \bibinfo {author} {\bibfnamefont {P.~H.}\ \bibnamefont {Fuoss}}, \bibinfo
  {author} {\bibfnamefont {G.~B.}\ \bibnamefont {Stephenson}}, \bibinfo
  {author} {\bibfnamefont {M.}~\bibnamefont {Cammarata}}, \bibinfo {author}
  {\bibfnamefont {D.~M.}\ \bibnamefont {Fritz}}, \bibinfo {author}
  {\bibfnamefont {A.}~\bibnamefont {Robert}},\ and\ \bibinfo {author}
  {\bibfnamefont {G.}~\bibnamefont {Gr\"ubel}},\ }\bibfield  {title} {\bibinfo
  {title} {Single shot spatial and temporal coherence properties of the slac
  linac coherent light source in the hard x-ray regime},\ }\href
  {https://doi.org/10.1103/PhysRevLett.108.024801} {\bibfield  {journal}
  {\bibinfo  {journal} {Phys. Rev. Lett.}\ }\textbf {\bibinfo {volume} {108}},\
  \bibinfo {pages} {024801} (\bibinfo {year} {2012})}\BibitemShut {NoStop}%
\bibitem [{\citenamefont {Sikorski}\ \emph {et~al.}(2016)\citenamefont
  {Sikorski}, \citenamefont {Feng}, \citenamefont {Song}, \citenamefont {Zhu},
  \citenamefont {Carini}, \citenamefont {Herrmann}, \citenamefont {Nishimura},
  \citenamefont {Hart},\ and\ \citenamefont
  {Robert}}]{sikorski2016application}%
  \BibitemOpen
  \bibfield  {author} {\bibinfo {author} {\bibfnamefont {M.}~\bibnamefont
  {Sikorski}}, \bibinfo {author} {\bibfnamefont {Y.}~\bibnamefont {Feng}},
  \bibinfo {author} {\bibfnamefont {S.}~\bibnamefont {Song}}, \bibinfo {author}
  {\bibfnamefont {D.}~\bibnamefont {Zhu}}, \bibinfo {author} {\bibfnamefont
  {G.}~\bibnamefont {Carini}}, \bibinfo {author} {\bibfnamefont
  {S.}~\bibnamefont {Herrmann}}, \bibinfo {author} {\bibfnamefont
  {K.}~\bibnamefont {Nishimura}}, \bibinfo {author} {\bibfnamefont
  {P.}~\bibnamefont {Hart}},\ and\ \bibinfo {author} {\bibfnamefont
  {A.}~\bibnamefont {Robert}},\ }\bibfield  {title} {\bibinfo {title}
  {Application of an epix100 detector for coherent scattering using a hard
  x-ray free-electron laser},\ }\href@noop {} {\bibfield  {journal} {\bibinfo
  {journal} {J. Synchrot. Radiat.}\ }\textbf {\bibinfo {volume} {23}},\
  \bibinfo {pages} {1171} (\bibinfo {year} {2016})}\BibitemShut {NoStop}%
\bibitem [{\citenamefont {Sun}\ \emph {et~al.}(2020{\natexlab{b}})\citenamefont
  {Sun}, \citenamefont {Montana-Lopez}, \citenamefont {Fuoss}, \citenamefont
  {Sutton},\ and\ \citenamefont {Zhu}}]{sun2020accurate}%
  \BibitemOpen
  \bibfield  {author} {\bibinfo {author} {\bibfnamefont {Y.}~\bibnamefont
  {Sun}}, \bibinfo {author} {\bibfnamefont {J.}~\bibnamefont {Montana-Lopez}},
  \bibinfo {author} {\bibfnamefont {P.}~\bibnamefont {Fuoss}}, \bibinfo
  {author} {\bibfnamefont {M.}~\bibnamefont {Sutton}},\ and\ \bibinfo {author}
  {\bibfnamefont {D.}~\bibnamefont {Zhu}},\ }\bibfield  {title} {\bibinfo
  {title} {Pulse intensity characterization of the lcls nanosecond double-bunch
  mode of operation},\ }\href@noop {} {\bibfield  {journal} {\bibinfo
  {journal} {J. Synchrotron Rad.}\ }\textbf {\bibinfo {volume} {27}},\ \bibinfo
  {pages} {999} (\bibinfo {year} {2020}{\natexlab{b}})}\BibitemShut {NoStop}%
\bibitem [{\citenamefont {Hruszkewycz}\ \emph {et~al.}(2012)\citenamefont
  {Hruszkewycz}, \citenamefont {Sutton}, \citenamefont {Fuoss}, \citenamefont
  {Adams}, \citenamefont {Rosenkranz}, \citenamefont {Ludwig}, \citenamefont
  {Roseker}, \citenamefont {Fritz}, \citenamefont {Cammarata}, \citenamefont
  {Zhu}, \citenamefont {Lee}, \citenamefont {Lemke}, \citenamefont {Gutt},
  \citenamefont {Robert}, \citenamefont {Gr\"ubel},\ and\ \citenamefont
  {Stephenson}}]{hruszkewycz2012high}%
  \BibitemOpen
  \bibfield  {author} {\bibinfo {author} {\bibfnamefont {S.~O.}\ \bibnamefont
  {Hruszkewycz}}, \bibinfo {author} {\bibfnamefont {M.}~\bibnamefont {Sutton}},
  \bibinfo {author} {\bibfnamefont {P.~H.}\ \bibnamefont {Fuoss}}, \bibinfo
  {author} {\bibfnamefont {B.}~\bibnamefont {Adams}}, \bibinfo {author}
  {\bibfnamefont {S.}~\bibnamefont {Rosenkranz}}, \bibinfo {author}
  {\bibfnamefont {K.~F.}\ \bibnamefont {Ludwig}}, \bibinfo {author}
  {\bibfnamefont {W.}~\bibnamefont {Roseker}}, \bibinfo {author} {\bibfnamefont
  {D.}~\bibnamefont {Fritz}}, \bibinfo {author} {\bibfnamefont
  {M.}~\bibnamefont {Cammarata}}, \bibinfo {author} {\bibfnamefont
  {D.}~\bibnamefont {Zhu}}, \bibinfo {author} {\bibfnamefont {S.}~\bibnamefont
  {Lee}}, \bibinfo {author} {\bibfnamefont {H.}~\bibnamefont {Lemke}}, \bibinfo
  {author} {\bibfnamefont {C.}~\bibnamefont {Gutt}}, \bibinfo {author}
  {\bibfnamefont {A.}~\bibnamefont {Robert}}, \bibinfo {author} {\bibfnamefont
  {G.}~\bibnamefont {Gr\"ubel}},\ and\ \bibinfo {author} {\bibfnamefont
  {G.~B.}\ \bibnamefont {Stephenson}},\ }\bibfield  {title} {\bibinfo {title}
  {High contrast x-ray speckle from atomic-scale order in liquids and
  glasses},\ }\href {https://doi.org/10.1103/PhysRevLett.109.185502} {\bibfield
   {journal} {\bibinfo  {journal} {Phys. Rev. Lett.}\ }\textbf {\bibinfo
  {volume} {109}},\ \bibinfo {pages} {185502} (\bibinfo {year}
  {2012})}\BibitemShut {NoStop}%
\bibitem [{\citenamefont {Roseker}\ \emph {et~al.}(2018)\citenamefont
  {Roseker}, \citenamefont {Hruszkewycz}, \citenamefont {Lehmk{\"u}hler},
  \citenamefont {Walther}, \citenamefont {Schulte-Schrepping}, \citenamefont
  {Lee}, \citenamefont {Osaka}, \citenamefont {Str{\"u}der}, \citenamefont
  {Hartmann}, \citenamefont {Sikorski} \emph {et~al.}}]{roseker2018towards}%
  \BibitemOpen
  \bibfield  {author} {\bibinfo {author} {\bibfnamefont {W.}~\bibnamefont
  {Roseker}}, \bibinfo {author} {\bibfnamefont {S.}~\bibnamefont
  {Hruszkewycz}}, \bibinfo {author} {\bibfnamefont {F.}~\bibnamefont
  {Lehmk{\"u}hler}}, \bibinfo {author} {\bibfnamefont {M.}~\bibnamefont
  {Walther}}, \bibinfo {author} {\bibfnamefont {H.}~\bibnamefont
  {Schulte-Schrepping}}, \bibinfo {author} {\bibfnamefont {S.}~\bibnamefont
  {Lee}}, \bibinfo {author} {\bibfnamefont {T.}~\bibnamefont {Osaka}}, \bibinfo
  {author} {\bibfnamefont {L.}~\bibnamefont {Str{\"u}der}}, \bibinfo {author}
  {\bibfnamefont {R.}~\bibnamefont {Hartmann}}, \bibinfo {author}
  {\bibfnamefont {M.}~\bibnamefont {Sikorski}}, \emph {et~al.},\ }\bibfield
  {title} {\bibinfo {title} {Towards ultrafast dynamics with split-pulse x-ray
  photon correlation spectroscopy at free electron laser sources},\ }\href@noop
  {} {\bibfield  {journal} {\bibinfo  {journal} {Nat. Commun.}\ }\textbf
  {\bibinfo {volume} {9}},\ \bibinfo {pages} {1704} (\bibinfo {year}
  {2018})}\BibitemShut {NoStop}%
\bibitem [{\citenamefont {Glownia}\ \emph {et~al.}(2019)\citenamefont
  {Glownia}, \citenamefont {Gumerlock}, \citenamefont {Lemke}, \citenamefont
  {Sato}, \citenamefont {Zhu},\ and\ \citenamefont
  {Chollet}}]{glownia2019pump}%
  \BibitemOpen
  \bibfield  {author} {\bibinfo {author} {\bibfnamefont {J.~M.}\ \bibnamefont
  {Glownia}}, \bibinfo {author} {\bibfnamefont {K.}~\bibnamefont {Gumerlock}},
  \bibinfo {author} {\bibfnamefont {H.~T.}\ \bibnamefont {Lemke}}, \bibinfo
  {author} {\bibfnamefont {T.}~\bibnamefont {Sato}}, \bibinfo {author}
  {\bibfnamefont {D.}~\bibnamefont {Zhu}},\ and\ \bibinfo {author}
  {\bibfnamefont {M.}~\bibnamefont {Chollet}},\ }\bibfield  {title} {\bibinfo
  {title} {Pump--probe experimental methodology at the linac coherent light
  source},\ }\href@noop {} {\bibfield  {journal} {\bibinfo  {journal} {Journal
  of synchrotron radiation}\ }\textbf {\bibinfo {volume} {26}},\ \bibinfo
  {pages} {685} (\bibinfo {year} {2019})}\BibitemShut {NoStop}%
\end{thebibliography}%


%

\end{document}